\documentclass[journal]{IEEEtran}

\usepackage{float}
\usepackage[table]{xcolor}
\usepackage{booktabs}
\usepackage{hyperref}
\usepackage{enumitem}
\usepackage{amsmath} 
\usepackage{amsfonts} 
\usepackage{amssymb}
\usepackage{multirow} 
\usepackage{booktabs}
\usepackage{graphicx}
\usepackage{stfloats}
\usepackage{tabularx}
\usepackage{CJKutf8}
\usepackage{xcolor}

\usepackage{amsmath,amsfonts}
\usepackage{algorithmic}
\usepackage{algorithm}
\usepackage{array}
\usepackage[caption=false,font=normalsize,labelfont=sf,textfont=sf]{subfig}
\usepackage{textcomp}
\usepackage{stfloats}
\usepackage{url}
\usepackage{verbatim}
\usepackage{graphicx}
\usepackage{cite}
\definecolor{groupgray}{RGB}{235,235,235}
\definecolor{lightblue}{RGB}{225,240,255}

\begin{document}

\title{To Fuse or to Drop? \\Dual-Path Learning for Resolving Modality Conflicts in Multimodal Emotion Recognition}

\author{Yangchen Yu, Qian Chen, Jia Li, Zhenzhen Hu, Jinpeng Hu, Lizi Liao, 
Erik Cambria,~\IEEEmembership{Fellow,~IEEE,} and Richang~Hong,~\IEEEmembership{Member,~IEEE} 

% \thanks{Yangchen Yu and Qian Chen contributed equally to this work.}
% \thanks{Corresponding author: Jia Li (e-mail: jiali@hfut.edu.cn).}
% \thanks{Yangchen Yu, Qian Chen, Jia Li, Zhenzhen Hu, Jinpeng Hu, and Richang Hong are with the School of Computer Science and Information Engineering, Hefei University of Technology, Hefei 230601, China.}
% \thanks{Lizi Liao is with the Singapore Management University, Singapore.}
% \thanks{Erik Cambria is with the Nanyang Technological University, Singapore, and also with the MIT Media Lab, USA.}
% }
\thanks{

		Yangchen Yu, Qian Chen, Jia Li, Zhenzhen Hu, Jinpeng Hu and Richang~Hong are with the School of Computer Science and Information Engineering, Hefei University of Technology, Hefei 230601, China (e-mail: shijie@mail.hfut.edu.cn; 2025170835@mail.hfut.edu.cn; jiali@hfut.edu.cn; huzhen.ice@gmail.com; 135858hjp@gmail.com; hongrc.hfut@gmail.com).

        % 廖老师
        Lizi Liao is with the School of Computing and Information Systems, Singapore Management University, Singapore (e-mail: lzliao@smu.edu.sg).

        %Erik
        Erik Cambria is with the College of Computing and Data Science, Nanyang Technological University, Singapore, and also with the MIT Media Lab, Cambridge, MA, USA (e-mail: cambria@ntu.edu.sg).

	    Yangchen Yu and Qian Chen contributed equally to this work.
        
		Corresponding author: Jia Li.
	}
}



% % column for its text to clear the IEEEpubid mark.

\maketitle

% \begin{abstract}
% Multimodal emotion recognition (MER) benefits from combining text, audio, and vision, yet standard fusion often fails when modalities conflict. Crucially, conflicts differ in resolvability: some disagreements reflect missing or weak cues and can be mitigated by cross-modal calibration, while others arise from intrinsically contradictory (\textit{e.g.}, sarcasm) or even misleading signals  where forced fusion amplifies errors. Recognizing this, we propose \textbf{D}ual-Path \textbf{C}onflict \textbf{R}esolution (\textbf{DCR}), a unified framework that learns when to fuse and when to drop modalities. Path I (Affective Fusion Distiller, AFD) performs reverse distillation from audio/visual teachers to a textual student using temporally weighted class evidence, improving fusion when alignment is beneficial. Path II (Affective Discernment Agent, ADA) formulates MER as a contextual bandit that selects among fusion and unimodal predictions based on a dual-view state and a calibration-aware reward, enabling robust inference under irreconcilable conflicts without per-modality reliability labels. Across four benchmarks, DCR consistently outperforms competitive MER baselines, and ablations show AFD and ADA are complementary\footnote{Codes and data are available at \url{https://anonymous.4open.science/r/DCR/}}.
% \end{abstract}

\begin{abstract}
Multimodal emotion recognition (MER) benefits from combining text, audio, and vision, yet standard fusion often fails when modalities conflict. Crucially, conflicts differ in resolvability: benign conflicts stem from missing, weak, or ambiguous cues and can be mitigated by cross-modal calibration, while severe conflicts arise from intrinsically contradictory (\textit{e.g.}, sarcasm) or misleading signals, for which forced fusion may amplify errors. Recognizing this, we propose \textbf{D}ual-Path \textbf{C}onflict \textbf{R}esolution (\textbf{DCR}), a unified framework that learns when to fuse and when to drop modalities. Path I (Affective Fusion Distiller, AFD) performs reverse distillation from audio/visual teachers to a textual student using temporally weighted class evidence, thereby enhancing representation-level calibration and improving fusion when alignment is beneficial. Path II (Affective Discernment Agent, ADA) formulates MER as a contextual bandit that selects among fusion and unimodal predictions based on a dual-view state and a calibration-aware reward, enabling decision-level arbitration under irreconcilable conflicts without requiring per-modality reliability labels. By taking into account the full multimodal context and coupling soft calibration with hard arbitration, DCR reconciles conflicts that can be aligned while bypassing misleading modalities when fusion is harmful. Across five benchmarks covering both dialogue-level and clip-level MER, DCR consistently outperforms competitive baselines or achieves highly competitive results. Further ablations, conflict-specific subset evaluation, and modality-selection analysis verify that AFD and ADA are complementary and jointly improve robust conflict-aware emotion recognition\footnote{Source code and models will be released at \url{https://github.com/MSA-LMC/DCR}}.
\end{abstract}


\section{Introduction}
% \IEEEPARstart{M}{ultimodal} Emotion Recognition (MER) is pivotal for advancing human–computer interaction, healthcare, and robotics~\cite{liu2017facial}, aiming to infer human affect by integrating linguistic, visual, and acoustic cues~\cite{liang2021attention, lv2021progressive}. These heterogeneous modalities are theoretically complementary: text conveys semantic intent, while audio and vision capture prosody and non‑verbal affective dynamics. However,in practice, modalities often disagree due to semantic inconsistency, noise, or misleading signals, resulting in modality conflicts that degrade performance~\cite{li2023decoupled}. Conventional fusion strategies, such as feature concatenation, frequently fail to reconcile these discrepancies and can even underperform strong unimodal baselines. To address this issue, recent work has explored cross‑modal distillation and representation refinement, but most approaches rely on text‑centric supervision~\cite{yun-etal-2024-telme} or uniform self‑distillation across modalities~\cite{10.1109/TMM.2023.3271019}, assuming that such conflicts can be resolved within a shared latent space.

\IEEEPARstart{M}{ultimodal} Emotion Recognition (MER) is pivotal for advancing human-computer interaction, healthcare, and robotics~\cite{liu2017facial}, aiming to infer human affect by integrating linguistic, visual, and acoustic cues~\cite{liang2021attention, lv2021progressive}. These heterogeneous modalities are theoretically complementary: text usually conveys semantic intent, while audio and vision capture prosody and non-verbal affective dynamics. However, in practice, modalities often disagree due to semantic inconsistency, noise, missing cues, or misleading signals, resulting in modality conflicts that degrade performance~\cite{li2023decoupled}. For example, neutral utterances may become affectively clear only with supportive prosody, whereas sarcastic expressions may exhibit a direct contradiction between literal semantics and vocal or facial cues. Conventional fusion strategies, such as feature concatenation, frequently fail to reconcile these discrepancies and can even underperform strong unimodal baselines, because they indiscriminately aggregate all modalities regardless of their reliability~\cite{li2023decoupled}. To address this issue, recent work has explored cross-modal distillation and representation refinement, but most approaches rely on text-centric supervision~\cite{yun-etal-2024-telme} or uniform self-distillation across modalities~\cite{10.1109/TMM.2023.3271019}, implicitly assuming that modality conflicts can always be resolved within a shared latent space. Such an assumption limits their ability to handle cases where certain modalities convey contradictory or misleading affective signals.

\begin{figure}[t]
  \centering
  \vspace{-0.2cm}
  \includegraphics[width=\columnwidth]{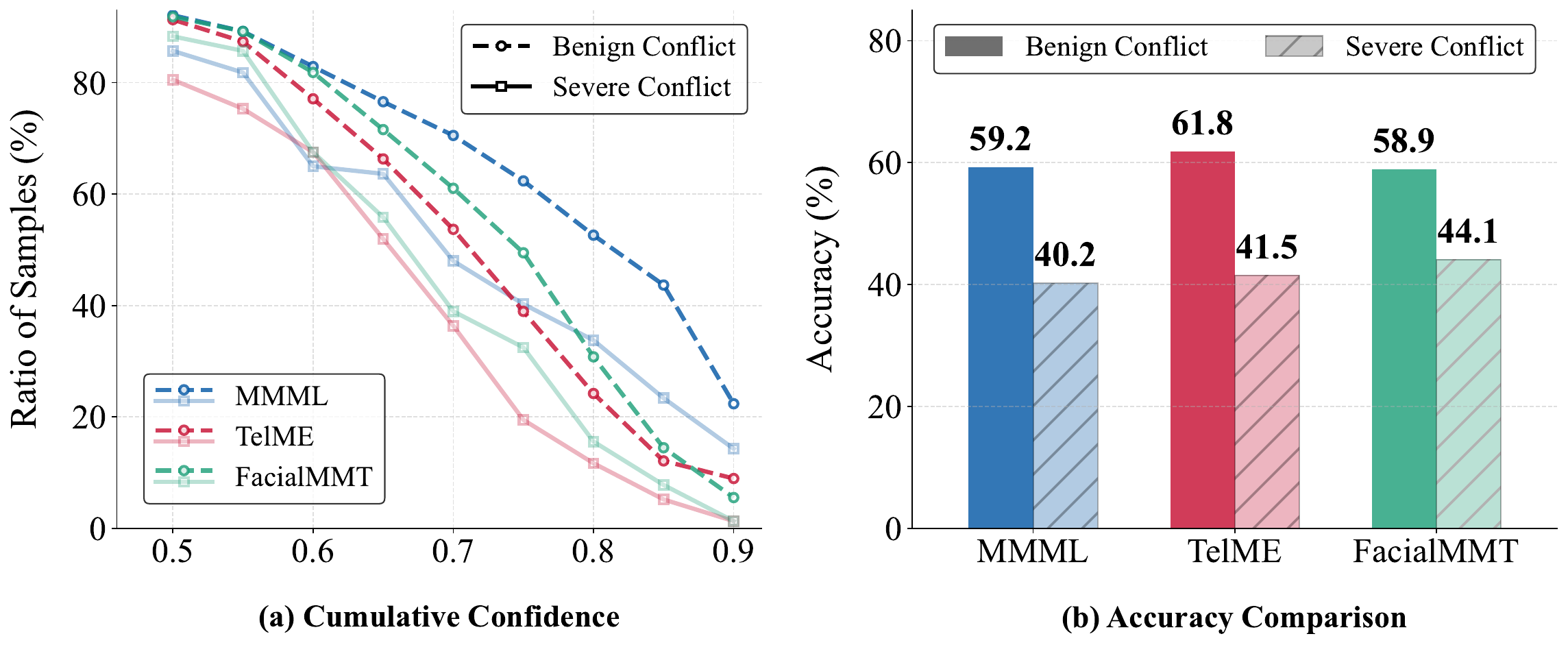} % 自动适配单栏宽度
  \caption{Empirical analysis of modality conflicts in common multimodal fusion for MER. The results illustrate the metrics of three representative and well-known fusion models, namely MMML~\cite{wu-etal-2024-multimodal}, TelME~\cite{yun-etal-2024-telme}, and FacialMMT~\cite{zheng-etal-2023-facial}, on the CH-SIMS~\cite{yu-etal-2020-ch} test dataset under different levels of modality conflicts. (a) Ratio of samples where the combined confidence of the top-$k$ (with $k=2$) predictions exceeds varying thresholds. (b) Accuracy comparison on benign versus severe modality conflict subsets.}
  \vspace{-0.2cm}
  \label{fig:motivation}
\end{figure}

% We challenge this assumption by viewing modality disagreements through the lens of resolvability. Concretely, we distinguish between benign (reconcilable) conflicts (e.g., neutral text refined by positive prosody) and severe (irreconcilable) conflicts (e.g., literal semantics inverted by sarcastic affect). Although the ideal definition of severity would measure the degradation per sample ‐ under an oracle MER model, this is infeasible in practice; therefore, we adopt a simple heuristic approximation for empirical tractability (Section~\ref{sec:def_con}). This abstraction provides the basis for the controlled simulations in Figure~\ref{fig:motivation}, enabling us to stress‑test conventional fusion models under different conflict intensity.

We revisit this assumption from the perspective of conflict resolvability. Instead of regarding all cross-modal discrepancies as errors to be aligned, we categorize modality conflicts into benign and severe cases. Benign conflicts involve weak or ambiguous evidence that can be refined by complementary modalities (e.g., neutral text refined by positive prosody), whereas severe conflicts involve contradictory or misleading evidence where forced fusion may amplify errors (e.g., literal semantics inverted by sarcastic affect). Although the ideal definition of severity would measure sample-level degradation under an oracle MER model, this is infeasible in practice; therefore, we adopt a simple heuristic approximation for empirical tractability (elaborated in Section~\ref{sec:def_con}). This abstraction provides the basis for the controlled simulations in Figure~\ref{fig:motivation}, enabling us to stress-test existing popular fusion models across conflict intensities and examine whether fusion remains reliable as conflicts become harder to reconcile.

The simulation results reveal the vulnerability of common multimodal fusion. As shown in Figure~\ref{fig:motivation}, existing models remain relatively confident under benign conflicts, yet both confidence and accuracy decline substantially as conflicts become irreconcilable, with the mean accuracy decreasing from 60\% to 42\%. These results indicate that indiscriminate fusion may amplify cross-modal interference when modalities provide inconsistent or misleading affective evidence. Therefore, modality conflicts should not be treated as a uniform alignment problem, but should be addressed according to their resolvability.

This distinction leads to two complementary principles. For benign conflicts, the model should calibrate the reliable textual branch with complementary non-verbal dynamics. Although text often provides stronger unimodal evidence, its affective meaning can vary across multimodal contexts~\cite{gan2024survey}, while acoustic and visual modalities capture temporal affective cues absent from transcripts. For severe conflicts, however, forced alignment may be counterproductive, as contradictory or misleading cues can introduce stronger interference. The model should therefore shift from representation-level alignment to decision-level arbitration, selecting the most reliable affective pathway when fusion becomes unreliable.

% For benign conflicts, the model should calibrate the reliable textual branch with complementary non-verbal dynamics. Although text often provides stronger unimodal evidence, its affective meaning can vary across multimodal contexts~\cite{gan2024survey}, while acoustic and visual modalities capture temporal affective cues absent from transcripts.

To this end, we propose the Dual-Path Conflict Resolution (DCR) framework, which handles modality disagreements according to their resolvability. Path I, the Affective Fusion Distiller (AFD), mitigates benign conflicts by distilling temporally weighted audio-visual affective evidence into the textual branch, enabling calibrated fusion when alignment is beneficial. Path II, the Affective Discernment Agent (ADA), addresses severe conflicts through policy-driven pathway selection, choosing among multimodal and unimodal predictions when fusion becomes unreliable. Together, DCR forms a soft-calibration-to-hard-arbitration framework that learns when to fuse complementary cues and when to bypass misleading modalities.

We summarize the primary contributions as follows:
\begin{itemize}[itemsep=0.1em, topsep=0.1em]
    % \item \textbf{Dual-Path Conflict Resolution (DCR).} For the first time, we propose a novel framework named DCR to resolve modality conflicts in MER according to their resolvability, reconciling benign conflicts while dynamically handling severe conflicts.

    % \item \textbf{Conflict-Aware Insight and Dual-Path Framework.} We introduce the insight that modality conflicts in MER vary in resolvability, and can be divided into benign and severe types. Building on this, we propose DCR, a unified dual-path framework that follows a divide-and-conquer paradigm: aligning modalities for benign conflicts while adaptively handling severe ones. This design provides a principled solution for conflict-aware multimodal emotion recognition.

    \item \textbf{Conflict-Aware Insight and Dual-Path Framework.} We identify that modality conflicts in MER differ in resolvability and divide them into benign and severe types. Based on this insight, we propose DCR, a unified dual-path framework that resolves them in a divide-and-conquer manner: aligning benign conflicts while adaptively handling severe ones.
    
    \item \textbf{Affective Fusion Distiller (AFD) and Affective Discernment Agent (ADA).} AFD mitigates benign conflicts via reverse knowledge distillation of non-verbal temporal dynamics, while ADA addresses severe conflicts through RL-based (policy-driven) modality selection.
    
    \item \textbf{Comprehensive analysis and SOTA performance.} DCR offers a novel principled analysis of benign versus severe conflicts and achieves SOTA performance on dialogue-level and clip-level MER benchmarks.
\end{itemize}

% \begin{figure*}[t]  % [t] 表示放置在页面顶部
%   \centering
%   \includegraphics[width=\textwidth]{samples.pdf} % 自动适配全页宽度
%   \caption{Representative samples of benign and severe conflict cases. $A, V, T$ and $M$ denote the ground-truth labels derived from unimodal (Acoustic, Visual, Textual) and multimodal annotation processes, respectively. Labels highlighted in \textbf{\color{red}red} indicate polarity inversion relative to the multimodal annotation $M$, intuitively reflecting the presence of severe modality conflict.}

%   \label{fig:samples}
% \end{figure*}

\section{Related Work}

\subsection{Multimodal Fusion for MER}

Multimodal emotion recognition (MER) aims to infer affective states by integrating complementary cues from language, vision, and acoustics. Existing studies have mainly advanced MER through multimodal fusion and cross-modal alignment~\cite{shou2026comprehensive}. Early methods focus on explicitly modeling inter-modal correlations, typically through tensor-based fusion or other algebraic interaction operations, so as to capture high-order relationships among heterogeneous modalities~\cite{zadeh-etal-2017-tensor, liu2018efficient, 10.5555/3504035.3504726}. These approaches demonstrate the effectiveness of multimodal integration, but they are often sensitive to noise, redundancy, and modality imbalance~\cite{gan2024survey}.

To alleviate these limitations, later studies introduce more structured representation learning strategies. A representative direction is to disentangle modality-invariant and modality-specific information, with the goal of preserving shared affective semantics while reducing irrelevant interference from individual modalities~\cite{hazarika2020misa, tdmtfer, 10.1145/3503161.3547754}. With the growing adoption of Transformer architectures, attention-based cross-modal interaction has further become a dominant paradigm in MER. For example, MulT~\cite{tsai-etal-2019-multimodal} employs directional cross-modal attention to model unaligned sequences, while PMR~\cite{Lv_2021_CVPR} progressively reinforces inter-modal interaction to improve semantic consistency. These methods significantly enhance the capacity of MER models to capture complex dependencies across modalities.

Despite such progress, most existing methods still follow a fusion-centric paradigm~\cite{gan2024survey} and implicitly assume that multimodal fusion is always beneficial. However, in realistic affective scenarios, different modalities may provide inconsistent, weak, or even contradictory emotional evidence~\cite{yu-etal-2020-ch}. Under these cases, directly enforcing multimodal interaction may not always resolve discrepancies, and can even amplify misleading signals~\cite{li2023decoupled}. Therefore, existing fusion methods provide only limited treatment of modality conflicts. In this work, we revisit MER from a conflict-aware perspective that jointly considers both multimodal fusion and unimodal inference.

\subsection{Beyond Conventional Fusion}

Beyond conventional fusion, recent studies have explored guided learning and adaptive decision mechanisms to improve robustness in multimodal systems. One important direction is knowledge distillation (KD), which transfers knowledge from a teacher model to a student model~\cite{Hinton2015DistillingTK}. KD has been extended from output-level logit matching to intermediate feature supervision and relational knowledge transfer~\cite{Romero2014FitNetsHF, chen2021reviewkd}. In multimodal learning, such strategies are often used to exploit complementary information across modalities, improve unimodal representations, and enhance cross-modal consistency~\cite{Zhang2025MultimodalKD}. These methods provide an effective way to inject additional supervision into multimodal representation learning.

In parallel, Reinforcement Learning (RL) has provided a principled framework for adaptive and discrete decision-making across various research fields, enabling models to dynamically select computation paths or input modalities~\cite{Han2021DynamicNN, Wang_2018_ECCV}. Recent studies have further applied RL to modality selection and reasoning under uncertainty, demonstrating its effectiveness in handling varying signal reliability~\cite{hu-etal-2024-mosel, Zhang2024RLEMOAR}.

However, these approaches have rarely been explored in MER from the perspective of affective semantics across modalities. Existing methods usually improve fusion quality, but they do not explicitly distinguish between conflicts that can be mitigated through cross-modal calibration and those that require avoiding harmful fusion. As a result, they lack a unified mechanism for deciding when to fuse and when to rely on more reliable unimodal pathways~\cite{muse2022,mer2024}. To address this gap, we explicitly model modality conflict as an inherent challenge in MER and propose a unified framework for conflict-aware resolution.

\begin{figure*}[t]  % [t] 表示放置在页面顶部
  \centering
  \includegraphics[width=0.96\textwidth]{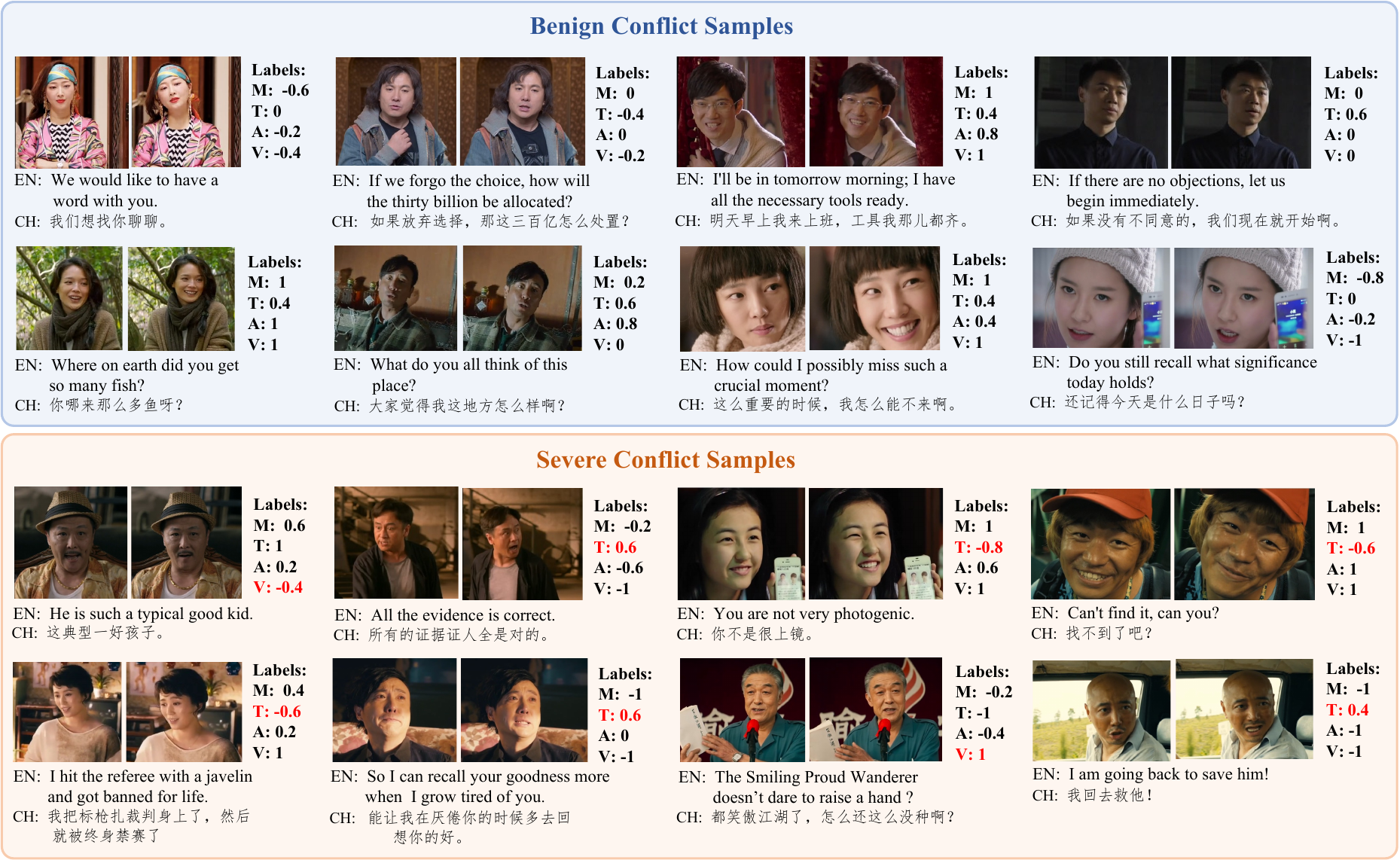} % 自动适配全页宽度
  \caption{Representative samples of benign and severe conflict cases under our heuristic conflict taxonomy. $A$, $V$, $T$, and $M$ denote the ground-truth labels from unimodal (Acoustic, Visual, Textual) and multimodal annotations, respectively. Labels highlighted in \textbf{\color{red}red} indicate polarity inversion relative to the multimodal annotation $M$, reflecting the presence of severe modality conflict.}

% Representative samples of benign and severe conflict cases under our heuristic conflict taxonomy. $A$, $V$, $T$, and $M$ denote the ground-truth labels from unimodal (Acoustic, Visual, Textual) and multimodal annotations, respectively. Labels highlighted in \textbf{\color{red}red} indicate polarity inversion relative to the multimodal annotation $M$, reflecting the presence of severe modality conflict.

  \label{fig:samples}
\end{figure*}

% 模型结构图
\begin{figure*}[t]  % [t] 表示放置在页面顶部
  \centering
  \includegraphics[width=0.96\textwidth]{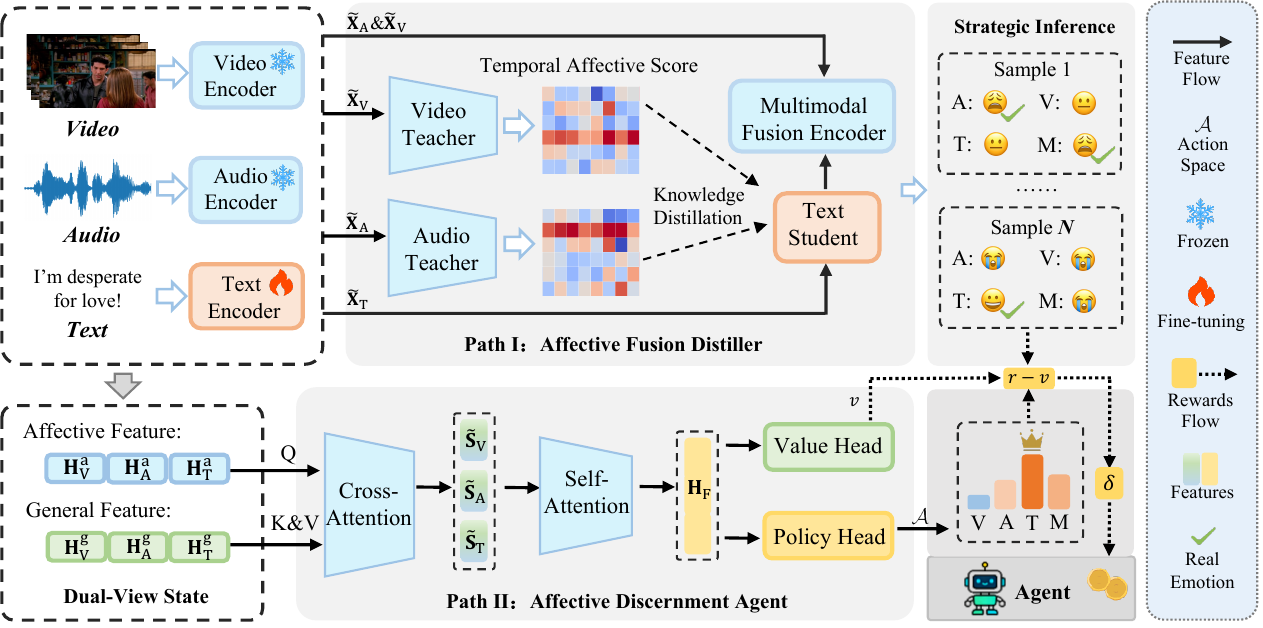} % 自动适配全页宽度
  \vspace{-0.1cm}
  \caption{The \textbf{DCR} framework consists of two specialized paths: (1) Path I (AFD), which reconciles modality conflicts via reverse knowledge distillation from frozen unimodal teachers to a text student for fusion-based prediction; and (2) Path II (ADA), which handles irreconcilable conflicts by employing a contextual bandit agent. Based on a dual-view state representation, the agent learns to select the most reliable modality or fusion result ($a \in \mathcal{A}$) and receives a reinforcement reward \textit{$\delta$} by checking against the optimal outcomes from Path I.}
  \vspace{-0.2cm}
  \label{fig:framework}
\end{figure*}

% \begin{figure*}[t]  % [t] 表示放置在页面顶部
%   \centering
%   \includegraphics[width=\textwidth]{samples.pdf} % 自动适配全页宽度
%   \caption{Representative samples of benign and severe conflict cases. $A, V, T$ and $M$ denote the ground-truth labels derived from unimodal (Acoustic, Visual, Textual) and multimodal annotation processes, respectively. Labels highlighted in \textbf{\color{red}red} indicate polarity inversion relative to the multimodal annotation $M$, intuitively reflecting the presence of severe modality conflict.}

%   \label{fig:samples}
% \end{figure*}

\section{Methodology}
\label{sec:method}

% As illustrated in Figure~\ref{fig:framework}, the DCR framework orchestrates two specialized pathways to resolve modality conflicts based on their resolvability. Our methodology follows a staged progression: Section~\ref{sec:afd} introduces the AFD for fine-grained knowledge transfer and soft calibration, while Section~\ref{sec:ada} describes the ADA’s role in decision-level hard arbitration. The technical description concludes in Section~\ref{sec:synergy} with their synergistic integration, enabling the model to transition from soft consensus to hard arbitration for robust affective inference.

As illustrated in Figure~\ref{fig:framework}, the DCR framework orchestrates two specialized pathways to resolve modality conflicts based on their resolvability. Our methodology proceeds in four stages. We begin by defining benign and severe conflicts in Section~\ref{sec:def_con}, thereby establishing the conceptual basis of the proposed framework. Section~\ref{sec:afd} then presents the Affective Fusion Distiller (AFD) for fine-grained knowledge transfer and soft calibration, while Section~\ref{sec:ada} introduces the Affective Discernment Agent (ADA) for decision-level hard arbitration. Section~\ref{sec:synergy} concludes the methodology by detailing their synergistic integration, through which the model transitions from soft calibration to hard arbitration for robust affective inference.

% 对良性以及硬性冲突进行概念的定义： 
\subsection{Benign and Severe Conflicts}
\label{sec:def_con}

% To ground our framework conceptually, we categorize modality conflicts into \textit{benign} and \textit{severe} according to their resolvability. Let $\mathbf{x}=\{x_T,x_A,x_V\}$ denote the multimodal input, where $T$, $A$, and $V$ correspond to textual, acoustic, and visual modalities, respectively. We consider an ideal Oracle MER model $\mathcal{F}^{*}$ that represents the upper bound of multimodal reasoning, and denote the ground-truth label as $y$. The severity of a modality conflict is defined by whether cross-modal disagreement alters the correct affective prediction:
% \begin{equation}
% \hat{y}^{*}_{M}=\mathcal{F}^{*}(x_T,x_A,x_V).
% \end{equation}
% A conflict is \textit{benign} if the Oracle prediction remains consistent with the ground truth:
% \begin{equation}
% \hat{y}^{*}_{M}=y,
% \end{equation}
% indicating that the disagreement is reconcilable through complementary cues. In contrast, a conflict is \textit{severe} if the disagreement leads to an incorrect prediction:
% \begin{equation}
% \hat{y}^{*}_{M}\neq y,
% \end{equation}
% suggesting that the modality discrepancy is irreconcilable under fusion (e.g., sarcasm where textual and nonverbal signals conflict).

To ground our framework conceptually, we categorize modality conflicts into \textit{benign} and \textit{severe} according to their resolvability. Let $\mathbf{x}=\{x_T,x_A,x_V\}$ denote the multimodal input, where $T$, $A$, and $V$ correspond to textual, acoustic, and visual modalities, respectively. We consider an ideal Oracle MER model $\mathcal{F}^{*}$ that represents the upper bound of multimodal reasoning, and denote the ground-truth affective target as $y$. The severity of a modality conflict is defined by whether cross-modal disagreement causes the Oracle prediction to deviate from $y$ beyond an acceptable tolerance:
\begin{equation}
\hat{y}^{*}_{M}=\mathcal{F}^{*}(x_T,x_A,x_V), 
\quad
s^{*}=\ell(\hat{y}^{*}_{M},y),
\end{equation}
% where $\ell(\cdot,\cdot)$ is a task-specific discrepancy function, and $\tau$ denotes a task-dependent tolerance threshold.
where $\ell(\cdot,\cdot)$ is a task-specific discrepancy function, such as misclassification error for discrete labels or absolute error for continuous affective scores, and $\tau$ denotes a task-dependent tolerance threshold.

A conflict is regarded as \textit{benign} if $s^{*}\leq \tau$, indicating that the disagreement remains reconcilable through complementary cues. In contrast, it is regarded as \textit{severe} if $s^{*}>\tau$, suggesting that the modality discrepancy is difficult to resolve through fusion (e.g., sarcasm where textual and nonverbal signals conflict).

Since such an Oracle is unavailable in practice, we adopt a heuristic approximation to make the taxonomy empirically operational. Rather than estimating the true causal effect of each modality, we use observable annotation patterns as a proxy for conflict severity. Specifically, we assess the consistency between unimodal polarities and the final multimodal judgment. A sample is considered severe if any unimodal polarity contradicts the multimodal polarity, and benign otherwise. Although this approximation is not a universal definition, it remains consistent with the conceptual definition above and provides a simple, reproducible way to construct conflict subsets for empirical analysis, as illustrated in Figure~\ref{fig:samples} and Figure~\ref{fig:motivation}.

\subsection{Path I: Affective Fusion Distiller}
\label{sec:afd}

% To cultivate a robust multimodal feature space, we introduce the Affective Fusion Distiller (AFD). Specifically, AFD targets benign conflicts by leveraging the rich temporal dynamics of non-verbal signals. By distilling these auxiliary cues, the module contextualizes static linguistic content into a nuanced emotional context, thereby fostering cross-modal synergy. 

To handle \textit{benign} conflicts defined above, we introduce the Affective Fusion Distiller (AFD), which leverages the rich temporal dynamics of non-verbal signals to calibrate and refine multimodal representations. Specifically, AFD targets benign conflicts by leveraging the rich temporal dynamics of non-verbal signals. By distilling these auxiliary cues, the module contextualizes static linguistic content into a nuanced emotional context, thereby fostering cross-modal synergy.

% \paragraph{Feature Extraction.}
% As illustrated in Figure~\ref{fig:framework}, given multimodal inputs consisting of text ($T$), audio ($A$) and visual ($V$) streams, we first extract raw features $\mathbf{X}_m \in \mathbb{R}^{L_m \times d_m}$ through pre-trained backbones, where $m \in \{T, A, V\}$). To reconcile discrepancies in temporal resolution and dimensionality, we employ independent 1D temporal convolutions to project all modalities into a unified latent space, producing aligned features $\tilde{\mathbf{X}}_m \in \mathbb{R}^{L \times d}$ for downstream distillation.

\vspace{0.5em}
\noindent \textbf{Reverse Affective Distillation.}
Starting with aligned multimodal features $\tilde{\mathbf{X}}_m \in \mathbb{R}^{L \times d}$ for each modality $ m \in \{V, A, T\}$, where $L$ and $d$ denote the sequence length and feature dimensionality respectively, projected from pre-trained backbones via standard 1D temporal convolutions, we introduce a Reverse Affective Distillation strategy to address the manifold collapse triggered by cross-modal distribution discrepancies. Unlike conventional benchmark-driven methods~\cite{yun-etal-2024-telme, 10.1109/TMM.2023.3271019}, we prioritize non-verbal modalities (Audio and Visual) as affective teachers due to their superior emotional granularity. In this framework, the text modality is reframed as a student (see Figure~\ref{fig:framework}), supervised to inherit the rich temporal dynamics typically latent in static linguistic representations.

Specifically, for each non-verbal modality $n \in \{A, V\}$, AFD first derives raw class activation maps $\mathbf{I}_n \in \mathbb{R}^{L \times C}$ via Grad-CAM~\cite{zhou2016learning} to quantify the contribution of each time step to the emotion categories, where $L$ and $C$ denote the sequence length and the number of emotion classes, respectively.  
By mapping these gradient-based importance scores, the model can pinpoint the pivotal moments that define specific emotions. Consequently, the localized prediction distribution $\mathbf{P}_{n, l, c}$ for category $c$ at time step $l$ is formulated as:
% The resulting temporal prediction distribution $\mathbf{P}_{n, l, c}$ at time step $l$ for category $c$ is obtained by:

%公式1
\begin{equation}
\mathbf{P}_{n, l, c} = \frac{\exp(\mathbf{I}_{n, l, c})}{\sum_{k=1}^{C} \exp(\mathbf{I}_{n, l, k})}.
\end{equation}
However, $\mathbf{P}_n$ only captures relative scores within each time step and treats all segments equally, lacking an explicit measure of temporal saliency. To refine the supervisory signal, we introduce Temporal Confidence Weighting, which derives a saliency weight $\mathbf{w}^n$ by normalizing the activation trajectory of the ground-truth label $y$ across the temporal dimension. For the $t$-th time step, the weight $\mathbf{w}_t^{n}$ is calculated as:
%公式2
\begin{equation}
\mathbf{w}_t^{n} = \frac{\exp(\mathbf{I}_{n, t, y})}{\sum_{j=1}^{L} \exp(\mathbf{I}_{n, j, y})}.
\end{equation}

Once high-confidence temporal cues are obtained from the teacher modalities, the textual student projects its features $\tilde{\mathbf{X}}_T$ into the affective classification space to generate the prediction distribution $\mathbf{P}_{\mathcal{S}}^t$. To bridge the cross-modal semantic gap, AFD transfers affective knowledge from both acoustic and visual teachers to the textual student through confidence-weighted distillation. The distillation objective is formulated as:
\begin{equation}
\mathcal{L}_{KL} = \sum_{n \in \{A, V\}} \sum_{t=1}^{L} \mathbf{w}_{t}^n \cdot \mathcal{D}_{KL} \big( \mathbf{P}_{n}^t \Vert \mathbf{P}_{\mathcal{S}}^t \big),
\end{equation}
where $\mathcal{D}_{KL}(\cdot \Vert \cdot)$ denotes the Kullback-Leibler divergence, and $\mathbf{w}_{t}^n$ is the temporal confidence weight derived from teacher modality $n$. This objective encourages the textual student to absorb reliable affective cues from non-verbal modalities while accounting for cross-modal discrepancies. Through this process, the model learns not only \textit{what} emotion is expressed but also \textit{when} it is expressed, enabling a more fine-grained understanding of temporal emotional dynamics.

\vspace{0.5em}
\noindent \textbf{Multi-Tier Synergistic Objective in Path I.}
To foster cross-modal synergy and ensure unimodal reliability, we optimize a comprehensive multi-tier joint objective. Specifically, a cross-attention mechanism is employed to dynamically align and aggregate modality-specific features into a unified fused representation. From this integrated feature space, the model generates a multimodal prediction, supervised by the multimodal classification loss $\mathcal{L}_{M}$. Meanwhile, to prevent over-reliance on a single modality and maintain independent discriminative power, we apply unimodal supervision to the textual, acoustic, and visual branches, where the corresponding unimodal classification loss is denoted as $\mathcal{L}_{U}$. Both $\mathcal{L}_{M}$ and $\mathcal{L}_{U}$ are implemented using cross-entropy loss. Together with the confidence-weighted distillation loss $\mathcal{L}_{KL}$, the final training objective is formulated as:
\begin{equation}
\mathcal{L}_{AFD} = \mathcal{L}_{M} + \gamma  \mathcal{L}_{U} + \lambda \mathcal{L}_{KL},
\end{equation}
where $\gamma$ and $\lambda$ are balancing coefficients for unimodal supervision and distillation regularization, respectively.

\subsection{Path II: Affective Discernment Agent}
\label{sec:ada}

% To complement the feature-level distillation of AFD, we introduce ADA to resolve severe cross-modal conflicts (e.g., irony) where conventional soft fusion often proves inadequate. We reformulate the selection of the optimal inference path as a \textbf{Contextual Multi-Armed Bandit (CMAB)} problem~\cite{langford2007epoch}. In this framework, ADA serves as a decision-level arbiter (Path II), dynamically adjudicating between ``synergistic fusion'' and ``unimodal selection'' to identify the most reliable affective path. By strategically filtering out misleading signals from conflicting modalities, ADA ensures robust inference even in complex conversational contexts.

% To address \textit{severe} conflicts, where cross-modal discrepancies are strong enough to overturn the correct affective judgment, we introduce the Affective Discernment Agent (ADA) as a decision-level arbitration module. We reformulate the selection of the optimal inference path as a \textbf{Contextual Multi-Armed Bandit (CMAB)} problem~\cite{langford2007epoch}. In this framework, ADA serves as a decision-level arbiter (Path II), dynamically adjudicating between ``synergistic fusion'' and ``unimodal selection'' to identify the most reliable affective path. By strategically filtering out misleading signals from conflicting modalities, ADA ensures robust inference even in complex conversational contexts.

To address \textit{severe} conflicts, where cross-modal discrepancies are strong enough to overturn the correct affective judgment, we introduce the Affective Discernment Agent (ADA) as a decision-level arbitration module. We reformulate the selection of the optimal inference path as a \textbf{Contextual Multi-Armed Bandit (CMAB)} problem~\cite{langford2007epoch}. In this framework, ADA serves as a decision-level arbiter (Path II), dynamically adjudicating between ``synergistic fusion'' and ``unimodal selection'' based on the joint multimodal context, thereby identifying the most reliable affective path. By strategically filtering out misleading signals from conflicting modalities, ADA ensures robust inference even in complex conversational contexts.

% To complement the feature-level calibration of AFD, ADA resolves severe conflicts (e.g., irony) where soft fusion proves inadequate. Formulated as a \textit{Contextual Multi-Armed Bandit (CMAB)}~\cite{langford2007epoch}, ADA serves as a decision-level arbiter, dynamically adjudicating between "synergistic fusion" and "unimodal selection" to identify the optimal affective inference path to filter out misleading signals.

\vspace{0.5em}
\noindent \textbf{State Construction and Augmentation.}
% To mitigate task-specific biases and ensure robust discernment, we construct a dual-view state space as the context for ADA, as shown in Figure~\ref{fig:framework}. For each modality $m$, the state integrates subjective affective features ($\mathbf{H}_{m}^a$) with objective general features ($\mathbf{H}_{m}^g$) derived from large-scale pre-trained backbones.While $\mathbf{H}_{m}^a$ captures specialized emotional nuances, it is prone to empirical overfitting. In contrast, $\mathbf{H}_{m}^g$ provides an impartial, robust context. This design compels ADA to "fact-check" task-specific cues against a broader perceptual environment, grounding its decisions in both specialized expertise and generalizable context.
As illustrated in Figure~\ref{fig:framework}, we construct a \textbf{Dual-view state space}\footnote{An empirical observation supporting this design is discussed in Section~\ref{sec:ablation study}, and its effectiveness is verified in Table~\ref{tab:component_ablation}.} to provide a reliable context for ADA. For each modality $m$, the state integrates two complementary components: \textit{subjective affective features} ($\mathbf{H}_{m}^a$) and \textit{objective general features} ($\mathbf{H}_{m}^g$). While $\mathbf{H}_{m}^a$ captures specific emotional cues from the current task, $\mathbf{H}_{m}^g$ offers a stable background derived from large-scale pre-training. This design helps ADA balance specialized emotional signals with a broader context, ensuring a more generalizable basis for its decisions.

% \footnote{An empirical observation supporting this design is discussed in Section~\ref{sec:sec:ablation study}, and its effectiveness is verified in Table~\ref{tab:component_ablation}.}

To further improve the model's robustness under cross-modal conflicts, we introduce a modality-level data augmentation strategy before feeding the state into the agent. Specifically, we apply stochastic modality dropout to each training sample, with probabilities $p_1$ and $p_2$ to mask one or two modalities, respectively. Additionally, Gaussian noise with standard deviation $\sigma$ is injected into the feature representations to simulate environmental uncertainty. This scheme encourages the agent to learn robust decision policies even when confronted with incomplete or misleading signals, enhancing its performance across diverse conversational scenes.

% [Note: Data Augmentation details to be inserted here as requested]
% 这里会有一个数据增强的内容，我想可以卸载环境构造的部分： 具体内容参见附录的什么位置

\vspace{0.5em}
\noindent \textbf{Cognitive Calibration and Discrete Decision.}
% To rectify empirical biases, ADA anchors subjective insights against objective context through Cognitive Calibration. Specifically, we employ a cross-attention mechanism using subjective features $\mathbf{H}_{m}^a$ as the Query ($\mathbf{Q}_m$) to retrieve supporting evidence from objective counterparts $\mathbf{H}_{m}^g$ (Key $\mathbf{K}_m$ and Value $\mathbf{V}_m$):
To rectify potential empirical biases, ADA anchors subjective affective insights against the objective context through Cognitive Calibration. This mechanism leverages the broad semantic knowledge embedded in general-purpose features to provide a stable reference for the task-specific emotional representations. Specifically, we employ a cross-attention mechanism where subjective affective features $\mathbf{H}_{m}^a$ act as the Query ($\mathbf{Q}_m$) to retrieve and integrate supporting evidence from the objective counterparts $\mathbf{H}_{m}^g$, which serve as the Key ($\mathbf{K}_m$) and Value ($\mathbf{V}_m$):

\begin{equation}
\tilde{\mathbf{S}}_m = \text{Softmax}\left( \frac{\mathbf{Q}_m \mathbf{K}_m^\top}{\sqrt{d_k}} \right) \mathbf{V}_m,
\end{equation}
where $d$ denotes the projection dimension and $\tilde{\mathbf{S}}_m \in \mathbb{R}^d$ represents the bias-rectified state for each modality $m$. To maintain modality-specific identity during the fusion process, these rectified states are concatenated with modality identity embeddings before being processed by a Transformer Encoder. This encoder captures the complex inter-modal dependencies and generates a unified global context representation $\mathbf{H}_M$. Finally, a policy head $\pi_\theta$, typically implemented as a multi-layer perceptron (MLP) with a Softmax activation, maps this global context to a probability distribution over the discrete action space $\mathcal{A} = \{a_M, a_T, a_A, a_V\}$, encompassing both the integrated multimodal path and independent unimodal branches: 

% where $d$ is the projection dimension and $\tilde{\mathbf{S}}_m$ denotes the bias-rectified state. By concatenating these states with modality identity embeddings, a Transformer Encoder generates the global context $\mathbf{H}_{M}$.
% Finally, a policy head $\pi_\theta$ map it to a probability distribution over the action space $\mathcal{A} = \{ a_{M}, a_T, a_A, a_V \}$, encompassing both the integrated multimodal path and independent unimodal branches:

\begin{equation}
\pi_\theta(a|\mathbf{H}_{M}) = \text{Softmax}(\text{MLP}_{policy}(\mathbf{H}_{M})).
\end{equation}
% This enables ADA to function as a dynamic routing agent: it can either exploit synergistic gains via $a_{M}$ or activate a unimodal path to target the real emotion (see Figure~\ref{fig:framework}), ensuring the most credible emotional inference.
% This enables ADA to function as a dynamic routing agent: it can either exploit synergistic gains via the integrated path $a_{M}$ or activate a specific unimodal path to isolate the genuine emotion (see Figure~\ref{fig:framework}). By strategically filtering out misleading signals from conflicting modalities, ADA ensures the most credible and robust affective inference.
This enables ADA to function as a dynamic routing agent: based on the joint multimodal context, it can either exploit synergistic gains via the integrated path $a_{M}$ or activate a specific unimodal path when certain modalities become unreliable (see Figure~\ref{fig:framework}). By making path selection from a holistic multimodal view, ADA strategically filters out misleading signals from conflicting modalities and ensures more credible and robust affective inference.

\vspace{0.5em}
\noindent \textbf{Policy-Driven Arbitration Optimization  in Path II.}
% To optimize the strategic path selection with reduced variance, we employ the Advantage Actor-Critic (A2C) framework~\cite{pmlr-v48-mniha16}. We design a Calibration-Aware Reward $r$ to steer the agent toward both accuracy and predictive calibration:
To optimize path selection while reducing gradient variance, we employ the Advantage Actor-Critic (A2C) framework~\cite{pmlr-v48-mniha16}. Firstly, we define a Calibration-Aware Reward $r$ to evaluate the immediate action of the policy network. This reward is scaled by the agent's confidence in its decision: a correct prediction yields a positive reward amplified by high confidence, whereas an incorrect choice incurs a penalty that increases with the degree of overconfidence. This initial weighting scheme compels ADA to prioritize inference paths that are not only accurate but also consistently trustworthy:

\begin{equation}
r = \underbrace{\mathbb{I}(\mathbf{\hat{y}} = \mathbf{y}) \cdot \mathbf{p}[\mathbf{y}]}_{\text{Positive Incentive}} - \underbrace{\mathbb{I}(\mathbf{\hat{y}} \neq \mathbf{y}) \cdot \mathbf{p}[\mathbf{\hat{y}}]}_{\text{Negative Penalty}},
\end{equation}
% where $\mathbf{y}$ and $\mathbf{\hat{y}}$ denote the ground-truth label and the predicted emotion category of the selected path, respectively. $\mathbb{I}(\cdot)$ is the indicator function, and $\mathbf{p}$ represents the prediction distribution. 
where $\mathbf{y}$ and $\mathbf{\hat{y}}$ denote the ground-truth label and the predicted emotion category of the selected path, respectively. $\mathbb{I}(\cdot)$ is the indicator function, and $\mathbf{p}$ represents the prediction distribution, which softens the binary correctness signal into a confidence-aware reward for more informative policy learning.

% As illustrated in Figure~\ref{fig:framework}, following the A2C paradigm, the advantage signal is derived as $\delta = r - v$ to reduce gradient variance, where $v$ is the state value estimated by value head. Consequently, the optimization objective for ADA during training is formulated as:
 % As illustrated in Figure~\ref{fig:framework}, we calculate the advantage signal $\delta = r - v$ to refine the policy update. This follows the A2C paradigm, where the actor (policy head) learns the optimal path selection while the critic (value head) estimates the state value $v$ to serve as a baseline. By subtracting this baseline from the reward, the advantage signal $\delta$ effectively reduces gradient variance and stabilizes the training process. 
 As illustrated in Figure~\ref{fig:framework}, we compute the advantage signal $\delta = r - v$, where $v$ denotes the state value estimated by the critic as a baseline for the current state. This follows the A2C paradigm, in which the actor (policy head) learns the optimal path selection while the critic (value head) evaluates the expected return of the state. By subtracting this baseline from the reward, the advantage signal $\delta$ reduces gradient variance and stabilizes the policy update. Consequently, the optimization objective for ADA is formulated as:
\begin{equation}
% \mathcal{L}_{ADA} = -\log \pi_\theta(a|\mathbf{H}_{F}) \mathbf{\delta} + \alpha \mathbf{\delta}^2 - \beta \mathcal{H}(\pi_\theta)
\mathcal{L}_{ADA} = \mathcal{L}_{pg} + \alpha \mathcal{L}_{val} - \beta \mathcal{H}(\pi_\theta),
\end{equation}
where $\mathcal{L}_{pg} = \mathbb{E}_{a \sim \pi_\theta} [-\log \pi_\theta(a|\mathbf{H}_{M}) \delta]$ denotes the policy gradient loss, encouraging actions with positive advantages to refine modality discernment strategies; $\mathcal{L}_{val} = \mathbb{E}[\delta^2]$ represents the value loss which constrains the value head to provide stable baseline estimates; and $\mathcal{H}(\pi_\theta) = \mathbb{E}_{a \sim \pi_\theta} [-\log \pi_\theta(a|\mathbf{H}_{M})]$ denotes the policy entropy regularizer conditioned on $\mathbf{H}_{M}$ to maintain sufficient exploration across the heterogeneous modality space. The hyperparameters $\alpha$ and $\beta$ serve as trade-off coefficients to balance these optimization objectives.

This objective compels ADA to identify the optimal affective inference path by maximizing predictive reliability.

\subsection{Synergistic Dual-Path Framework}
\label{sec:synergy}

% The DCR framework addresses diverse multimodal conflicts by synergizing soft calibration with hard selection. Specifically, we introduce a Sequential Dependency paradigm, where both pathways leverage their mutual strengths to resolve discrepancies across dual dimensions.
The DCR framework addresses diverse multimodal conflicts by synergizing soft calibration with hard selection. We implement a hierarchical and complementary Sequential Dependency paradigm, where both pathways leverage their mutual strengths to resolve discrepancies across different granularity levels. This multi-stage architecture ensures that the model can adaptively handle a wide spectrum of conflicts, ranging from subtle feature noise to severe modality contradictions.

\vspace{0.5em}
\noindent \textbf{Path I: Feature-Level Calibration (AFD).} 
The synergy begins with AFD, which addresses \textit{benign} conflicts through feature-level soft calibration. Instead of simple feature alignment, AFD distills the temporal emotional changes from non-verbal modalities into the textual space. This process enhances the fine-grained perception of both unimodal and multimodal features, resolving minor discrepancies and stabilizing the representations. Consequently, AFD outputs four refined predictions including one multimodal and three unimodal results to serve as reliable candidates for the next stage.

\vspace{0.5em}
\noindent \textbf{Path II: Decision-Level Arbitration (ADA).} 
% Based on these calibrated predictions, ADA then resolves \textit{severe} conflicts through decision-level hard arbitration. When modalities contradict each other (e.g., irony), ADA acts as a functional arbiter to select the most trustworthy inference path. It chooses from four candidates: the fused multimodal result or one of the three independent unimodal sources. This selection allows the model to bypass misleading information, ensuring robust emotional inference in complex scenarios.
Based on the calibrated modality representations and predictions, ADA then resolves \textit{severe} conflicts through decision-level hard arbitration. When modalities contradict each other (e.g., irony), ADA evaluates the joint multimodal context and cross-modal semantic consistency to determine which pathway provides more reliable affective evidence. It selects from four candidate channels: the multimodal fusion path or one of the three unimodal paths. This path-level arbitration allows the model to reduce the influence of misleading modalities and route inference through more credible affective cues, ensuring robust emotion recognition in complex scenarios.

In summary, by combining representation-level refinement with decision-level selection, the joint optimization objective for the DCR framework is formulated as:

% The core of this synergy lies in their hierarchical complementarity: \textbf{Path I (AFD)} resolves benign conflicts through representation-level calibration, thereby establishing a stabilized, high-quality action space $\mathcal{A}$ for subsequent decision-making. Conversely, \textbf{Path II (ADA)} introduces a hard selection mechanism to navigate severe conflicts that transcend the limits of soft calibration. By transitioning from soft consensus to hard arbitration, this staged architecture integrates the specific strengths of both pathways to ensure robust and credible emotional inference. This coordinated process in training phase is formally constrained as:

% \begin{gather}
%     \hat{\Theta}_{Agent} = \arg\min_{\Theta_{Agent}} \mathcal{L}_{ADA} \left( \pi_\theta, \mathcal{A}(\hat{\Theta}_{AFD}) \right) \\
%     \text{s.t.} \quad \hat{\Theta}_{AFD} = \arg\min_{\Theta_{AFD}} \mathcal{L}_{AFD},
% \end{gather}

\begin{gather}
\hat{\Theta}_{Agent}
= \arg\min_{\Theta_{Agent}}
\mathcal{L}_{ADA}\!\left(
\pi_\theta, \mathcal{A}(\hat{\Theta}_{AFD})
\right) \notag \\
\text{s.t.}\quad
\hat{\Theta}_{AFD}
= \arg\min_{\Theta_{AFD}} \mathcal{L}_{AFD},
\end{gather}
where the optimal policy $\hat{\Theta}_{Agent}$ is learned over the fixed action space $\mathcal{A}$ spanned by the optimized experts $\hat{\Theta}_{AFD}$. 

This staggered optimization ensures that the agent learns a robust decision policy over a group of well-calibrated experts. By sequentially aligning features and then arbitrating conflicts, the framework achieves stable and credible multimodal inference on emotion.

\section{Experiments}

\subsection{Datasets}
% We evaluate DCR on four widely used MER benchmarks, including two dialogue-level datasets (MELD and CMU-MOSEI) and two clip-level datasets (CH-SIMS and CH-SIMS v2). \textbf{MELD}~\cite{poria-etal-2019-meld}, derived from the TV series \textit{Friends}, is a multi-party conversational emotion recognition dataset containing 13,708 utterances annotated with seven emotion categories. We follow its official split of 9,989 training samples, 1,109 validation samples, and 2,610 test samples. \textbf{CMU-MOSEI}~\cite{bagher-zadeh-etal-2018-multimodal} contains 22,856 YouTube video clips annotated with sentiment intensity scores ranging from -3 to +3, covering seven levels from highly negative to highly positive. Following the standard evaluation protocol, we use 16,326 samples for training, 1,871 for validation, and 4,659 for testing. \textbf{CH-SIMS}~\cite{yu-etal-2020-ch} is a Chinese multimodal benchmark with 2,281 video clips, each annotated with multimodal labels, unimodal labels, and sentiment intensity scores in the range of -1 to +1. \textbf{CH-SIMS v2}~\cite{10.1145/3536221.3556630} extends CH-SIMS by including 4,403 labeled clips and 10,161 unlabeled clips; in our experiments, we use only the labeled portion. Detailed statistics and evaluation metrics of all datasets are summarized in Table~\ref{tab:dataset_statistics_metrics}.

We evaluate DCR on five widely used MER benchmarks, including three \textbf{\emph{dialogue-level datasets}} (MELD, IEMOCAP, and CMU-MOSEI) and two \textbf{\emph{clip-level datasets}} (CH-SIMS and CH-SIMS v2). \textbf{MELD}~\cite{poria-etal-2019-meld}, derived from the TV series \textit{Friends}, is a multi-party conversational emotion recognition dataset containing 13,708 utterances annotated with seven emotion categories. We follow its official split of 9,989 training samples, 1,109 validation samples, and 2,610 test samples. \textbf{IEMOCAP}~\cite{busso2008iemocap} is a multimodal dataset consisting of dyadic conversations between ten professional actors. It contains 151 dialogues with 7,433 utterances annotated with six emotion categories, including Neutral, Happy, Sad, Angry, Frustrated, and Excited. Specifically, the dataset includes 5,163 utterances for training, 647 utterances for validation, and 1,623 utterances for testing. \textbf{CMU-MOSEI}~\cite{bagher-zadeh-etal-2018-multimodal} contains 22,856 YouTube video clips annotated with sentiment intensity scores ranging from -3 to +3, covering seven levels from highly negative to highly positive. Following the standard evaluation protocol, we use 16,326 samples for training, 1,871 for validation, and 4,659 for testing. \textbf{CH-SIMS}~\cite{yu-etal-2020-ch} is a Chinese multimodal benchmark with 2,281 video clips, each annotated with multimodal labels, unimodal labels, and sentiment intensity scores in the range of -1 to +1. \textbf{CH-SIMS v2}~\cite{10.1145/3536221.3556630} extends CH-SIMS by including 4,403 labeled clips and 10,161 unlabeled clips; in our experiments, we use only the labeled portion. Detailed statistics and evaluation metrics of all datasets are summarized in Table~\ref{tab:dataset_statistics_metrics}.

\begin{table}[H]
\centering
\renewcommand{\arraystretch}{1.1}
\captionsetup{font=small}
\small
\caption{Statistics and evaluation metrics of the datasets.}
\label{tab:dataset_statistics_metrics}
\begin{tabular*}{\columnwidth}{@{\extracolsep{\fill}}lcccc} % 修改为5列
\specialrule{1.0pt}{0pt}{2pt}
\textbf{Dataset} & \textbf{Train} & \textbf{Valid} & \textbf{Test} & \textbf{Evaluation Metrics} \\
\midrule
MELD       & 9,989  & 1,109 & 2,610 & Acc-7, WF1  \\
IEMOCAP    & 5,163   & 647  & 1,623  & Acc-6, WF1  \\
CMU-MOSEI  & 16,326 & 1,871 & 4,659 & F1, Acc-7, MAE, Corr \\
CH-SIMS    & 1,368  & 456   & 457   & Acc-2, MAE \\
CH-SIMS V2 & 2,645  & 879   & 879   & Acc-2, MAE \\
\specialrule{1.0pt}{2pt}{0pt}
\end{tabular*}
\vspace{-0.3cm}
% \caption{Statistics and evaluation metrics of the datasets.}
% \label{tab:dataset_statistics_metrics}
\end{table}

\subsection{Baselines}
We compare DCR with representative MER baselines spanning several mainstream paradigms: Recurrent-based methods (e.g., DialogueCRN~\cite{hu2021dialoguecrn}, DialogueRNN~\cite{2019DialogueRNN}, and SACL-LSTM~\cite{hu-etal-2023-supervised}), which emphasize contextual temporal modeling in conversational emotion recognition; Transformer-based methods (e.g., MulT~\cite{tsai-etal-2019-multimodal}, Self-MM~\cite{yu2021le}, UniMSE~\cite{hu-etal-2022-unimse}, FacialMMT~\cite{zheng-etal-2023-facial}, SDT~\cite{10.1109/TMM.2023.3271019}, TelME~\cite{yun-etal-2024-telme} and RMER-DT~\cite{ZHU2025103268}), which leverage attention mechanisms to model multimodal interactions. Graph-based methods (e.g., MMGCN~\cite{hu-etal-2021-mmgcn}, GraphCFC~\cite{li2023graphcfc}, and ECERC~\cite{zhang-tan-2025-ecerc}) capture relational dependencies among speakers or across modalities using graph structures.
Additionally, we include competitive models tailored to specific benchmarks: for CH-SIMS, we consider LMF~\cite{liu-etal-2018-efficient-low}, HFR-AME~\cite{HFRAME} and DashFusion~\cite{11040049}; for MELD, we include BC-LSTM~\cite{poria2017context}, Joyful~\cite{li-etal-2023-joyful}, DialogueGCN~\cite{ghosal2019dialoguegcn} and DQ-Former~\cite{11480881}; for CMU-MOSEI, we evaluate ConKI~\cite{yu-etal-2023-conki}, CLGSI~\cite{yang-etal-2024-clgsi}, KEBR~\cite{10.1145/3664647.3681163}, MAG-BERT~\cite{rahman-etal-2020-integrating}, and MFON~\cite{zhang-etal-2025-modal}. 

We further include controlled baselines under the same backbone setting as DCR, including simple concatenation-based fusion and vanilla cross-attention fusion. The corresponding results are reported together with the ablation comparisons in Table~\ref{tab:modality_ablation}.

\subsection{Implementation Details}
% \section{Implementation Details}
\label{sec:implement details}

\vspace{0.5em}
\noindent \textbf{Evaluation Metrics.}
% We follow the standard evaluation protocols for each benchmark. For MELD, we report seven-class accuracy (Acc-7) and weighted F1 score (WF1), and additionally provide per-class accuracy for fine-grained analysis. For CMU-MOSEI, we report F1 score, Acc-7, Pearson correlation (Corr), and mean absolute error (MAE). Following prior work, the F1 score on CMU-MOSEI is presented in two forms using the ``-/-'' split marker, where the first value corresponds to negative versus non-negative classification and the second to negative versus positive classification. For CH-SIMS and CH-SIMS v2, we report binary accuracy (Acc-2) and MAE. Detailed dataset statistics and the corresponding evaluation metrics are summarized in Table~\ref{tab:dataset_statistics_metrics}. Among these metrics, lower values indicate better performance for MAE, while higher values are preferred for all the others.
We follow the standard evaluation protocols for each benchmark. For MELD, we report seven-class accuracy (Acc-7) and Weighted F1 score (WF1), and additionally provide per-class accuracy for fine-grained analysis. For IEMOCAP, we report six-class accuracy (Acc-6) and Weighted F1 score (WF1) following common practice in conversational emotion recognition. For CMU-MOSEI, we report F1 score, Acc-7, Pearson correlation (Corr), and Mean Absolute Error (MAE). Following prior work, the F1 score on CMU-MOSEI is presented in two forms using the ``-/-'' split marker, where the first value corresponds to negative versus non-negative classification and the second to negative versus positive classification. For CH-SIMS and CH-SIMS v2, we report binary accuracy (Acc-2) and Mean Absolute Error (MAE). Detailed dataset statistics and the corresponding evaluation metrics are summarized in Table~\ref{tab:dataset_statistics_metrics}. Among these metrics, lower values indicate better performance for MAE, while higher values are preferred for all the others.

\vspace{0.5em}
\noindent \textbf{Experimental Setup.}
All models are implemented in PyTorch and trained on a single NVIDIA RTX 4090 GPU with 24GB memory. To reduce the effect of stochastic initialization, we conduct experiments with three to five random seeds selected from \{41, 42, 43, 44, 45\}, and report the averaged results. For feature extraction, we adopt RoBERTa-large as the linguistic encoder, Whisper-large-v3 as the acoustic encoder, and CLIP-ViT-B/16 as the visual encoder. Instead of using pooled global representations, we extract sequential features from all pretrained backbones to preserve fine-grained temporal dynamics.

For  \emph{dialogue-level} datasets, i.e., MELD, IEMOCAP and CMU-MOSEI, the entire conversation is fed into the textual encoder to capture contextual dependencies, followed by utterance-level emotion inference. In contrast, for \emph{clip-level} datasets, i.e., CH-SIMS and CH-SIMS v2, the transcript of each clip is processed independently. For modality-level data augmentation in ADA, we set the dropout probabilities to $p_1=0.2$ and $p_2=0.05$ for masking one and two modalities, respectively, and use Gaussian noise with $\sigma=0.01$. Unless otherwise specified, all models are trained using the Adam optimizer with a learning rate of $1\times10^{-4}$ and a batch size of 32.

\begin{table*}[t]
    \centering
    \small
    \captionsetup{font=small}
    \setlength{\tabcolsep}{10pt}
    \caption{Comparison on the MELD dialogue-level MER benchmark. \textbf{Bold} indicates the best result.}
    \label{tab:meld}
    \begin{tabular}{lccccccccc}
        \specialrule{1.1pt}{0pt}{2pt}
        Methods
        & Neutral & Surprise & Fear & Sadness & Joy & Disgust & Anger & Acc-7 & WF1 \\
        \midrule

        \rowcolor{groupgray}
        \multicolumn{10}{l}{\textit{Recurrent-based}} \\
        DialogueRNN~\cite{2019DialogueRNN}
        & 82.17 & 46.62 & - & 21.15 & 49.50 & - & 48.41 & 60.27 & 57.95 \\
        BC-LSTM~\cite{poria2017context}
        &- &- &- &- &- &- &- & 65.87 & 64.87 \\
        EmoCaps~\cite{li-etal-2022-emocaps}
        & 77.12 & 63.19 & 3.03 & 42.52 & 57.50 & 7.69 & 57.54 & - & 64.00 \\
        DialogueCRN~\cite{hu2021dialoguecrn}
        & 79.72 & 57.62 & 18.26 & 39.30 & 64.56 & 32.07 & 52.53 & 66.93 & 65.77 \\
        SACL-LSTM~\cite{hu-etal-2023-supervised}
        & 80.17 & 58.77 & 26.23 & 41.34 & 64.98 & 31.47 & 52.35 & 67.51 & 66.45 \\

        \midrule
        \rowcolor{groupgray}
        \multicolumn{10}{l}{\textit{Graph-based}} \\
        DialogueGCN~\cite{ghosal2019dialoguegcn}
        & - &- &- &- & -& -& - & 63.62 & 62.68 \\
        MMGCN~\cite{hu-etal-2021-mmgcn}
        & 84.32 & 47.33 & 2.00 & 14.90 & 56.97 & 1.47 & 42.61 & 61.34 & 58.41 \\
        GraphCFC~\cite{li2023graphcfc}
        & 76.98 & 49.36 & - &  26.89 & 51.88 & 47.59 & - &  61.42 & 58.86 \\
        Joyful~\cite{li-etal-2023-joyful}
        & 76.80 & 51.91 & - & 41.78 & 56.89 & - & 50.71 & 62.53 & 61.77 \\
        ECERC~\cite{zhang-tan-2025-ecerc}
        & 79.80 & 58.98 & 26.12 & 40.95 & 64.95 & 31.43 & 53.89 & 66.46 & 67.32 \\

        \midrule
        \rowcolor{groupgray}
        \multicolumn{10}{l}{\textit{Transformer-based}} \\
        % DQ-Former~\cite{11480881}
        % & 80.05 & 61.72 & 38.15 & 54.60 & 19.51 & 22.22 & 60.71 & 65.77 & 66.97 \\
        UniMSE~\cite{hu-etal-2022-unimse}
        & - & - & - & - & - & - & - & 65.09 & 65.51 \\
        FacialMMT~\cite{zheng-etal-2023-facial}
        & 80.13 & 59.63 & 19.18 & 41.99 & 64.88 & 18.18 & 56.00 & - & 66.58 \\
        SDT~\cite{10.1109/TMM.2023.3271019}
        & 83.22 & 61.28 & 13.80 & 34.90 & 63.24 & 22.65 & 56.93 & 67.55 & 66.60 \\
        % SACL-LSTM~\cite{hu-etal-2023-supervised}
        % & 80.17 & 58.77 & 26.23 & 41.34 & 64.98 & 31.47 & 52.35 & 67.51 & 66.45 \\
        DQ-Former~\cite{11480881}
        & 80.05 & 60.71 & 19.51 & 38.15 & 61.72 & 22.22 & 54.60 & 65.77 & 66.97 \\
        % EmoVerse~\cite{LI2025130810}
        % & 80.41 & 58.18 & \textbf{29.73} & 44.88 & 63.61 & \textbf{27.72} & 53.80 & 67.78 & 66.74 \\
        RMER-DT~\cite{ZHU2025103268}
        & 81.55 & 63.70 & 25.15 & \textbf{52.65} & \textbf{73.25} & 25.95 & \textbf{57.85} & - & 67.02 \\
        TelME~\cite{yun-etal-2024-telme}
        & 80.22 & 60.33 & 26.97 & 43.45 & 65.67 & 26.42 & 56.70 & - & 67.37 \\

        % \midrule
        \rowcolor{lightblue}
        \textbf{DCR (ours)}
        & \textbf{85.43}
        & \textbf{66.19}
        & 14.00
        & 42.31
        & 64.18
        & 25.00
        & 55.94
        & \textbf{69.81}
        & \textbf{68.84} \\

        \specialrule{1.1pt}{2pt}{3pt}
    \end{tabular}
    \vspace{-0.1cm}
\end{table*}

% IEMOCAP Results
\begin{table}[t]
    \centering
    \captionsetup{font=small}
    \small
    \setlength{\tabcolsep}{8pt}
    \caption{Comparison on the IEMOCAP benchmark. \textbf{Bold} indicates the best result.}
    \label{tab:iemocap}
    \begin{tabular}{lcc}
        \specialrule{1.1pt}{0pt}{2pt}
        Methods & Acc-6 & WF1 \\
        \midrule
        DialogueRNN~\cite{2019DialogueRNN} 
        & 64.85 & 64.65 \\
        MMGCN~\cite{hu-etal-2021-mmgcn} 
        & 65.80 & 65.73 \\
        DialogueCRN~\cite{hu2021dialoguecrn} 
        & 67.39 & 67.54 \\
        
        GraphCFC~\cite{li2023graphcfc} 
        & 69.13 & 68.91 \\
        SACL-LSTM~\cite{hu-etal-2023-supervised} 
        & 69.08 & 69.22 \\
        \midrule
        \rowcolor{lightblue}
        \textbf{DCR (ours)}
        & \textbf{69.85}
        & \textbf{69.50} \\
        \specialrule{1.1pt}{2pt}{3pt}
    \end{tabular}
\end{table}

% \begin{table}[t]
%     \centering
%     \captionsetup{font=small}
%     \small
%     \setlength{\tabcolsep}{8pt}
%     \caption{Comparison on IEMOCAP. Best results are highlighted in \textbf{bold}.}
%     \label{tab:iemocap}
%     \begin{tabular}{lcc}
%         \specialrule{1.1pt}{0pt}{2pt}
%         \textbf{Methods} & Acc-6 & WF1  \\
%         \midrule
%         % BC-LSTM~\cite{} &   63.08    &  62.84  \\
%         % DialogueCRN~\cite{hu2021dialoguecrn} &  67.39    &   67.54      \\
%         DialogueRNN~\cite{2019DialogueRNN} & 64.85 & 64.65 \\
%         MMGCN~\cite{hu-etal-2021-mmgcn} & 65.80 & 65.73 \\
%         DialogueCRN~\cite{hu2021dialoguecrn} &  67.39    &   67.54      \\
%         SDT~\cite{10.1109/TMM.2023.3271019} & 68.47 & 68.78 \\
%         GraphCFC~\cite{li2023graphcfc} & 69.13 & 68.91 \\
%         SACL-LSTM~\cite{hu-etal-2023-supervised} & 69.08 & 69.22 \\
%         \midrule
%         \rowcolor{lightblue}
%         \textbf{DCR (ours)}   & \textbf{69.85}  &  \textbf{69.50} \\
%         \specialrule{1.1pt}{2pt}{3pt}
%     \end{tabular}
% \end{table}

% CMU-MOSEI Results
\begin{table}[t]
    \centering
    \captionsetup{font=small}
    \small
    \setlength{\tabcolsep}{3.5pt}
    \caption{Comparison on the CMU-MOSEI benchmark. \textbf{Bold} indicates the best result.}
    \label{tab:mosei}
    \begin{tabular}{lcccc}
        \specialrule{1.1pt}{0pt}{2pt}
        Methods & F1 & Acc-7 & Corr & MAE $(\downarrow)$ \\
        \midrule
        MulT~\cite{tsai-etal-2019-multimodal} 
        & 81.05/83.46 & 52.34 & 0.600 & 0.671 \\
        MAG-BERT~\cite{rahman-etal-2020-integrating} 
        & 82.77/84.71 & 52.67 & 0.755 & 0.543 \\
        Self-MM~\cite{yu2021le} 
        & 82.95/84.93 & 53.46 & 0.767 & 0.529 \\
        ConKI~\cite{yu-etal-2023-conki} 
        & 83.08/86.15 & 54.25 & 0.752 & 0.529 \\
        CLGSI~\cite{yang-etal-2024-clgsi} 
        & 84.21/86.18 & \textbf{54.56} & 0.763 & 0.532 \\
        KEBR~\cite{10.1145/3664647.3681163} 
        & 84.25/86.68 & 54.37 & 0.799 & 0.517 \\
        MFON~\cite{zhang-etal-2025-modal} 
        & 83.13/86.29 & 53.72 & 0.780 & 0.528 \\
        \midrule
        \rowcolor{lightblue}
        \textbf{DCR (ours)}
        & \textbf{84.62}/\textbf{87.19}
        & 54.26
        & \textbf{0.807}
        & \textbf{0.510} \\
        \specialrule{1.1pt}{2pt}{3pt}
    \end{tabular}
\end{table}
% \begin{table}[t]
%     \centering
%     \captionsetup{font=small}
%     \setlength{\tabcolsep}{4pt}
%     \small
%     \caption{Comparison on the CMU-MOSEI. \textbf{Bold} indicates the best result.}
%     % \vspace{-0.2cm}
%     \label{tab:mosei}
%     \resizebox{\columnwidth}{!}{
%         \begin{tabular}{lccccc}
%             \specialrule{1.1pt}{0pt}{2pt}
%             Methods & F1 & \text{Acc-7} & Corr & MAE ($\downarrow$) \\
%             \midrule
%             MulT~\cite{tsai-etal-2019-multimodal}       & 81.05/83.46 & 52.34 & 0.60  & 0.671 \\
%             MAG-BERT~\cite{rahman-etal-2020-integrating}  & 82.77/84.71 & 52.67 & 0.755 & 0.543 \\
%             Self-MM~\cite{yu2021le}   & 82.95/84.93 & 53.46 & 0.767 & 0.529 \\
%             ConKI~\cite{yu-etal-2023-conki}     & 83.08/86.15 & 54.25 & 0.752 & 0.529 \\
%             CLGSI~\cite{yang-etal-2024-clgsi}     & 84.21/86.18 & \textbf{54.56} & 0.763 & 0.532 \\
%             % EUAR~\cite{10.1145/3664647.3680949}      & 82.02/85.46 & 51.07 & 0.785 & 0.596 \\
%             KEBR~\cite{10.1145/3664647.3681163}      & 84.25/86.68 & 54.37 & 0.799 & 0.517 \\
%             MFON~\cite{zhang-etal-2025-modal}      & 83.13/86.29 & 53.72 & 0.780 & 0.528 \\
%             \midrule
%             \rowcolor{lightblue}
%             \textbf{DCR (ours)} & \textbf{84.62}/\textbf{87.19} & 54.26 & \textbf{0.807} & \textbf{0.510} \\
%             \specialrule{1.1pt}{2pt}{3pt}
%         \end{tabular}
%     }
% \end{table}

% CH-SIMS Results
\begin{table}[t]
    \centering
    \captionsetup{font=small}
    \small
    \setlength{\tabcolsep}{3.5pt}
    \caption{Performance comparison on the Chinese clip-level MER benchmarks. The best results are highlighted in \textbf{bold}.}
    \label{tab:chsims}
    \begin{tabular}{lcccc}
        \specialrule{1.1pt}{0pt}{2pt}
        \multirow{2}{*}{Methods} & \multicolumn{2}{c}{CH-SIMS} & \multicolumn{2}{c}{CH-SIMS v2} \\
        \cmidrule(lr){2-3} \cmidrule(lr){4-5}
        & Acc-2 & MAE $(\downarrow)$ & Acc-2 & MAE $(\downarrow)$ \\
        \midrule
        LMF~\cite{liu-etal-2018-efficient-low} & 77.77 & 0.441 & 74.18 & 0.367 \\
        MulT~\cite{tsai-etal-2019-multimodal} & 78.56 & 0.432 & 80.68 & 0.291 \\
        Self-MM~\cite{yu2021le} & 80.04 & 0.425 & 79.69 & 0.311 \\
        DashFusion~\cite{11040049} & 79.21 & 0.416 & - & - \\
        HFR-AME~\cite{HFRAME} & 80.50 & 0.412 & - & - \\
        \midrule
        \rowcolor{lightblue}
        \textbf{DCR (ours)} & \textbf{80.96} & \textbf{0.332} & \textbf{81.91} & \textbf{0.290} \\
        \specialrule{1.1pt}{2pt}{3pt}
    \end{tabular}
    \vspace{-0.1cm}
\end{table}

\subsection{Comparison with the State of the Art}

To comprehensively evaluate the effectiveness of DCR, we compare it with representative MER baselines on both dialogue-level and clip-level benchmarks, including MELD, IEMOCAP, CMU-MOSEI, CH-SIMS, and CH-SIMS v2. The quantitative comparisons are summarized in Tables~\ref{tab:meld}--\ref{tab:chsims}. Overall, DCR demonstrates strong and consistent performance across datasets with different annotation granularities, label spaces, and evaluation protocols.

On the dialogue-level MELD benchmark, DCR achieves the best overall performance, reaching 69.81\% in Acc-7 and 68.84\% in WF1, as shown in Table~\ref{tab:meld}. Compared with strong recent baselines such as SDT~\cite{10.1109/TMM.2023.3271019}, TelME~\cite{yun-etal-2024-telme}, ECERC~\cite{zhang-tan-2025-ecerc}, EmoVerse~\cite{LI2025130810}, and RMER-DT~\cite{ZHU2025103268}, DCR establishes a new state of the art in both overall metrics. In particular, DCR improves WF1 by 1.47 points over TelME~\cite{yun-etal-2024-telme} and by 1.82 points over RMER-DT~\cite{ZHU2025103268}, indicating that the proposed framework is effective in handling complex conversational emotions under multi-party settings. At the class level, DCR also achieves the best accuracy on several major categories, including \textit{neutral} and \textit{surprise}, with 85.43 and 66.19, respectively. These gains suggest that DCR is particularly effective in capturing dominant conversational affect while maintaining robust discrimination under emotionally ambiguous dialogue contexts.

On the IEMOCAP benchmark, DCR also achieves strong performance, as reported in Table~\ref{tab:iemocap}. Specifically, DCR obtains 69.85\% in Acc-6 and 69.50\% in WF1, outperforming all compared baselines on both metrics. Compared with the strongest baseline SACL-LSTM~\cite{hu-etal-2023-supervised}, DCR improves Acc-6 by 0.77 points and WF1 by 0.28 points. It also surpasses GraphCFC~\cite{li2023graphcfc} by 0.72 points in Acc-6 and 0.59 points in WF1. These results indicate that DCR remains effective in dyadic conversational emotion recognition and generalizes well to different dialogue-level MER settings.

On the CMU-MOSEI benchmark, Table~\ref{tab:mosei} shows that DCR achieves the best F1 score, the highest correlation, and the lowest MAE among all compared methods. Specifically, DCR reaches 84.62/87.19 in F1, 0.807 in Corr, and 0.510 in MAE, surpassing strong baselines such as KEBR~\cite{10.1145/3664647.3681163}, MFON~\cite{zhang-etal-2025-modal}, and CLGSI~\cite{yang-etal-2024-clgsi} on most metrics. Although its Acc-7 is slightly below the best reported value, DCR remains highly competitive on this metric while showing clear advantages on the regression-oriented criteria. This result is particularly meaningful for CMU-MOSEI, where sentiment prediction requires not only correct polarity classification but also accurate estimation of sentiment intensity. The consistent gains in Corr and MAE indicate that DCR can better preserve fine-grained affective information and produce more reliable continuous sentiment predictions.

On the clip-level Chinese benchmarks CH-SIMS and CH-SIMS v2, DCR consistently achieves the best results on all reported metrics, as summarized in Table~\ref{tab:chsims}. On CH-SIMS, DCR obtains 80.96\% in Acc-2 and 0.332 in MAE, outperforming the strongest baseline HFR-AME~\cite{HFRAME} by 0.46 points in Acc-2 and reducing MAE by a substantial margin of 0.080. On CH-SIMS v2, DCR further reaches 81.91\% Acc-2 and 0.290 MAE, exceeding MulT~\cite{tsai-etal-2019-multimodal} and Self-MM~\cite{yu2021le} on both metrics. These results demonstrate that DCR generalizes effectively to clip-level affective prediction in Chinese videos and remains robust under different data scales and annotation settings.

Taken together, the above results show that DCR yields consistently strong performance across both dialogue-level and clip-level MER benchmarks. The improvements on MELD and IEMOCAP highlight its effectiveness in conversational emotion recognition, the gains on CMU-MOSEI demonstrate its strength in fine-grained sentiment regression, and the results on CH-SIMS and CH-SIMS v2 verify its robustness in clip-level multimodal affective analysis. This overall pattern suggests that the proposed Dual-Path framework provides a broadly effective solution for multimodal emotion recognition under diverse evaluation scenarios.

% 消融实验表1： 两条路径的消融
% Ablation study: dual-path components
\begin{table}[t]
    \centering
    \captionsetup{font=small}
    \small
    \setlength{\tabcolsep}{5pt}
    \caption{Ablation study of the Dual-Path components in DCR on MELD and CMU-MOSEI. \textbf{Bold} indicates the best result.}
    \label{tab:path_ablation_final}
    \begin{tabular}{cc|cc|cc}
        \specialrule{1.1pt}{0pt}{2pt}
        \multirow{2}{*}{AFD} 
        & \multirow{2}{*}{ADA} 
        & \multicolumn{2}{c|}{MELD} 
        & \multicolumn{2}{c}{CMU-MOSEI} \\
        \cmidrule(lr){3-4} \cmidrule(lr){5-6}
        & & Acc-7 & WF1 & Acc-7 & MAE $(\downarrow)$ \\
        \midrule
        \rowcolor{lightblue}
        $\checkmark$ & $\checkmark$ 
        & \textbf{69.81} & \textbf{68.84} 
        & \textbf{54.26} & \textbf{0.510} \\
        $\checkmark$ & 
        & 68.93 & 68.12 
        & 52.13 & 0.547 \\
         & $\checkmark$ 
        & 69.00 & 68.29 
        & 53.18 & 0.518 \\
         & 
        & 68.73 & 67.26 
        & 51.48 & 0.555 \\
        \specialrule{1.1pt}{2pt}{3pt}
    \end{tabular}
\end{table}

\subsection{Ablation Study}
\label{sec:ablation study}

To evaluate the contribution of each key component within the DCR framework, we conduct systematic ablation experiments by comparing the full model against several variants. The quantitative results are summarized in Tables~\ref{tab:path_ablation_final}-\ref{tab:component_ablation}.

\vspace{0.5em}
\noindent \textbf{Effectiveness of Dual-Path Architecture.} 
To rigorously validate the contribution of each component, we compare the full DCR framework with its single-path variants and a vanilla cross-attention fusion baseline. As shown in Table~\ref{tab:path_ablation_final}, the baseline without both AFD and ADA corresponds to a standard cross-attention-based multimodal fusion model, which performs prediction without either soft calibration or hard arbitration.
Both the AFD-only and ADA-only variants consistently outperform this baseline, indicating that each path independently contributes to mitigating modality conflicts. Specifically, the AFD-only variant improves WF1 on MELD from 67.26\% to 68.12\%, demonstrating the benefit of representation-level soft calibration. Meanwhile, the ADA-only variant further raises WF1 to 68.29\%, highlighting the advantage of decision-level path selection in identifying more reliable affective cues.
Notably, the full DCR framework achieves the best performance among all variants. This result suggests that AFD and ADA are not merely additive, but complementary: AFD refines the feature space by alleviating benign conflicts, while ADA provides flexible decision-level arbitration for severe conflicts. Their combination enables DCR to handle modality conflicts more effectively across different scenarios.

% % 消融实验表1： 两条路径的消融
% % Ablation study: dual-path components
% \begin{table}[t]
%     \centering
%     \captionsetup{font=small}
%     \small
%     \setlength{\tabcolsep}{5pt}
%     \caption{Ablation study of the Dual-Path components in DCR on MELD and CMU-MOSEI. \textbf{Bold} indicates the best result.}
%     \label{tab:path_ablation_final}
%     \begin{tabular}{cc|cc|cc}
%         \specialrule{1.1pt}{0pt}{2pt}
%         \multirow{2}{*}{AFD} 
%         & \multirow{2}{*}{ADA} 
%         & \multicolumn{2}{c|}{MELD} 
%         & \multicolumn{2}{c}{CMU-MOSEI} \\
%         \cmidrule(lr){3-4} \cmidrule(lr){5-6}
%         & & Acc-7 & WF1 & Acc-7 & MAE $(\downarrow)$ \\
%         \midrule
%         \rowcolor{lightblue}
%         $\checkmark$ & $\checkmark$ 
%         & \textbf{69.81} & \textbf{68.84} 
%         & \textbf{54.26} & \textbf{0.510} \\
%         $\checkmark$ & 
%         & 68.93 & 68.12 
%         & 52.13 & 0.547 \\
%          & $\checkmark$ 
%         & 69.00 & 68.29 
%         & 53.18 & 0.518 \\
%          & 
%         & 68.73 & 67.26 
%         & 51.48 & 0.555 \\
%         \specialrule{1.1pt}{2pt}{3pt}
%     \end{tabular}
% \end{table}

% 消融实验表2： 对于模态的消融分析：
% Ablation study: modality analysis
\begin{table}[t]
    \centering
    \captionsetup{font=small}
    \small
    \setlength{\tabcolsep}{5pt}
    \caption{Performance comparison of unimodal experts and fusion baselines on MELD and CMU-MOSEI. ``Concat'' and ``Cross-Attention'' denote feature-level concatenation and vanilla cross-modal attention fusion baselines, respectively. \textbf{Bold} indicates the best result.}
    \label{tab:modality_ablation}
    \begin{tabular}{lcccc}
        \specialrule{1.1pt}{0pt}{2pt}
        \multirow{2}{*}{Model} 
        & \multicolumn{2}{c}{MELD} 
        & \multicolumn{2}{c}{CMU-MOSEI} \\
        \cmidrule(lr){2-3} \cmidrule(lr){4-5}
        & Acc-7 & WF1 & Acc-7 & MAE $(\downarrow)$ \\
        \midrule
        Text-Only  
        & 68.71 & 67.00 & 51.89 & 0.567 \\
        Audio-Only 
        & 57.93 & 53.64 & 48.67 & 0.594 \\
        Video-Only 
        & 48.14 & 40.33 & 40.47 & 0.802 \\
        \midrule
        Concat 
        & 66.01 & 64.45 & 50.01 & 0.581 \\
        Cross-Attention 
        & 68.73 & 67.26 & 51.48 & 0.555 \\
        \rowcolor{lightblue}
        \textbf{AFD}
        & \textbf{68.93} & \textbf{68.12}
        & \textbf{52.13} & \textbf{0.547} \\
        \specialrule{1.1pt}{2pt}{3pt}
    \end{tabular}
\end{table}

% 消融实验表3：对于部分小组件的消融分析
% Ablation study: internal components of ADA
\begin{table}[t]
    \centering
    \captionsetup{font=small}
    \small
    \setlength{\tabcolsep}{5pt}
    \caption{Ablation study of ADA components on MELD and CMU-MOSEI. Gen., Emo., Calib., and Aug. denote General, Emotion, Calibration-Aware, and Data Augmentation, respectively. \textbf{Bold} indicates the best result.}
    \label{tab:component_ablation}
    \begin{tabular}{lcccc}
        \specialrule{1.1pt}{0pt}{2pt}
        \multirow{2}{*}{Variant} & \multicolumn{2}{c}{MELD} & \multicolumn{2}{c}{MOSEI} \\
        \cmidrule(lr){2-3} \cmidrule(lr){4-5}
        & Acc-7 & WF1 & Acc-7 & MAE ($\downarrow$) \\
        \midrule
        \rowcolor{lightblue}
        \textbf{Full DCR} 
        & \textbf{69.81} & \textbf{68.84} & \textbf{54.26} & \textbf{0.510} \\
        \midrule
        w/o Gen. Feat. 
        & 68.58 & 67.76 & 52.9 & 0.533 \\
        w/o Emo. Feat. 
        & 68.74 & 66.11 & 53.1 & 0.556 \\
        w/o Calib. Reward 
        & 69.27 & 68.35 & 53.72 & 0.529 \\
        w/o Value Head 
        & 69.62 & 68.70 & 54.02 & 0.515 \\
        w/o Aug. 
        & 69.27 & 68.44 & 53.91 & 0.521 \\
        \specialrule{1.1pt}{2pt}{3pt}
    \end{tabular}
\end{table}

% % 消融实验表4：对于ADA可扩展的消融分析
% \begin{table}[t]
% \centering
% \caption{Performance comparison of individual unimodal experts and the proposed DCR framework on MELD and MOSEI datasets. ``Concat'' denotes the conventional feature-level fusion baseline, and bold results indicate the best performance.}
% \label{tab:modality_ablation}
% \resizebox{\columnwidth}{!}{
% \begin{tabular}{lcccc}
% \toprule
% \multirow{2}{*}{Methods} & \multicolumn{2}{c}{MELD} & \multicolumn{2}{c}{MOSEI} \\
% \cmidrule(lr){2-3} \cmidrule(lr){4-5}
% & Acc7 ($\uparrow$) & WF1 ($\uparrow$) & Acc7 ($\uparrow$) & MAE ($\downarrow$) \\
% \midrule
% TelME(标注)  & -- & -- & -- & -- \\
% TelME (w/ADA) & -- & -- & -- & -- \\
% \textbf{DCR (Ours) }& \textbf{69.81} & \textbf{68.84} & \textbf{54.26} & \textbf{0.510} \\
% \bottomrule
% \end{tabular}
% }
% \end{table}

\vspace{0.5em}

\noindent \textbf{Modality Analysis and the Impact of AFD.} 
% We analyze the individual contribution of each modality to evaluate the distillation strategy within AFD. As shown in Table~\ref{tab:modality_ablation}, the textual modality emerges as the strongest unimodal predictor. Notably, on the dialogue-based MELD dataset, the text-only expert achieves a 67.00\% WF1, significantly outperforming the audio (53.64\%) and video (40.33\%) modalities. 
% However, a standard "Concat" fusion actually underperforms the standalone text expert, with the WF1 dropping to 64.45\% on MELD and the MAE increasing to 0.581 on MOSEI. This experimental evidence clearly illustrates the "fusion dilemma," where simply combining non-verbal features degrades the performance of the dominant textual modality. AFD effectively resolves this issue by distilling temporal emotional dynamics into the textual representations. By refining these features, AFD not only mitigates the fusion-induced degradation but also surpasses the best unimodal performance, achieving 68.12\% WF1 on MELD and 0.547 MAE on MOSEI. These results validate that AFD successfully harvests cross-modal gains while maintaining the reliability of the textual foundation.
We further analyze the contribution of each modality to evaluate the distillation strategy in AFD. As shown in Table~\ref{tab:modality_ablation}, the textual modality is the strongest unimodal predictor, achieving 67.00\% WF1 on MELD and outperforming the audio-only (53.64\%) and video-only (40.33\%) variants. A similar trend is observed on CMU-MOSEI, where the text-only model also yields the best unimodal performance.
However, conventional fusion does not necessarily improve over the dominant textual modality. \textit{Concat} reduces performance to 64.45\% WF1 on MELD and 0.581 MAE on CMU-MOSEI. We further include a vanilla cross-attention fusion baseline as a stronger controlled comparison. Although it performs better than \textit{Concat}, it still remains inferior to AFD. In contrast, AFD achieves the best results on both datasets, reaching 68.12\% WF1 on MELD and 0.547 MAE on CMU-MOSEI. These results show that AFD can better exploit cross-modal information while preserving the reliability of the textual foundation.

% % 样例分析
% \begin{figure*}[t]  % [t] 表示放置在页面顶部
%   \centering
%   \includegraphics[width=\textwidth]{samples.pdf} % 自动适配全页宽度
%   \caption{Representative samples of benign and severe conflict cases. $A, V, T$ and $M$ denote the ground-truth labels derived from unimodal (Acoustic, Visual, Textual) and multimodal annotation processes, respectively. Labels highlighted in \textbf{\color{red}red} indicate polarity inversion relative to the multimodal annotation $M$, intuitively reflecting the presence of severe modality conflict.}

%   \label{fig:samples}
% \end{figure*}

\vspace{0.5em}

\noindent \textbf{Fine-Grained Analysis of ADA.}
We conduct a series of ablation experiments on MELD and CMU-MOSEI to evaluate how different internal components contribute to the policy-driven arbitration of ADA. As shown in Table~\ref{tab:component_ablation}, the full DCR model achieves the best performance on both datasets, reaching 69.81\% Acc-7 and 68.84\% WF1 on MELD, and 54.26\% Acc-7 and 0.510 MAE on CMU-MOSEI. The results indicate that the dual-view state space, which integrates both subjective emotional features ($H_m^a$) and objective general features ($H_m^g$), is a fundamental element for informed decision-making. Specifically, removing emotional features causes a clear performance drop on MELD, reducing WF1 from 68.84\% to 66.11\%, and also degrades CMU-MOSEI from 0.510 to 0.556 in MAE. This suggests that specialized affective cues are critical for the agent to perceive nuanced emotional states. Removing general features also leads to consistent degradation on both datasets, indicating that objective pretrained representations provide useful contextual support for more stable arbitration.

The reinforcement learning components further support the training process and policy stability. When the calibration-aware reward is removed, the performance drops to 68.35\% WF1 on MELD and 0.529 MAE on CMU-MOSEI, showing that confidence-aware feedback helps the agent distinguish between reliable and misleading modality paths. Similarly, excluding the value head reduces performance on both datasets, which points to the role of stable value estimation in policy optimization. Furthermore, disabling data augmentation also leads to consistent degradation, with WF1 decreasing to 68.44\% on MELD and MAE increasing to 0.521 on CMU-MOSEI. In our framework, this augmentation randomly masks modal information and injects perturbations to simulate incomplete or noisy inputs, encouraging the agent to learn a more flexible arbitration policy. Collectively, these results demonstrate that ADA functions as an adaptive arbitrator rather than a static selector, effectively suppressing misleading cues while preserving useful cross-modal gains.

\begin{figure}[t]
  \centering
  \includegraphics[width=\columnwidth]{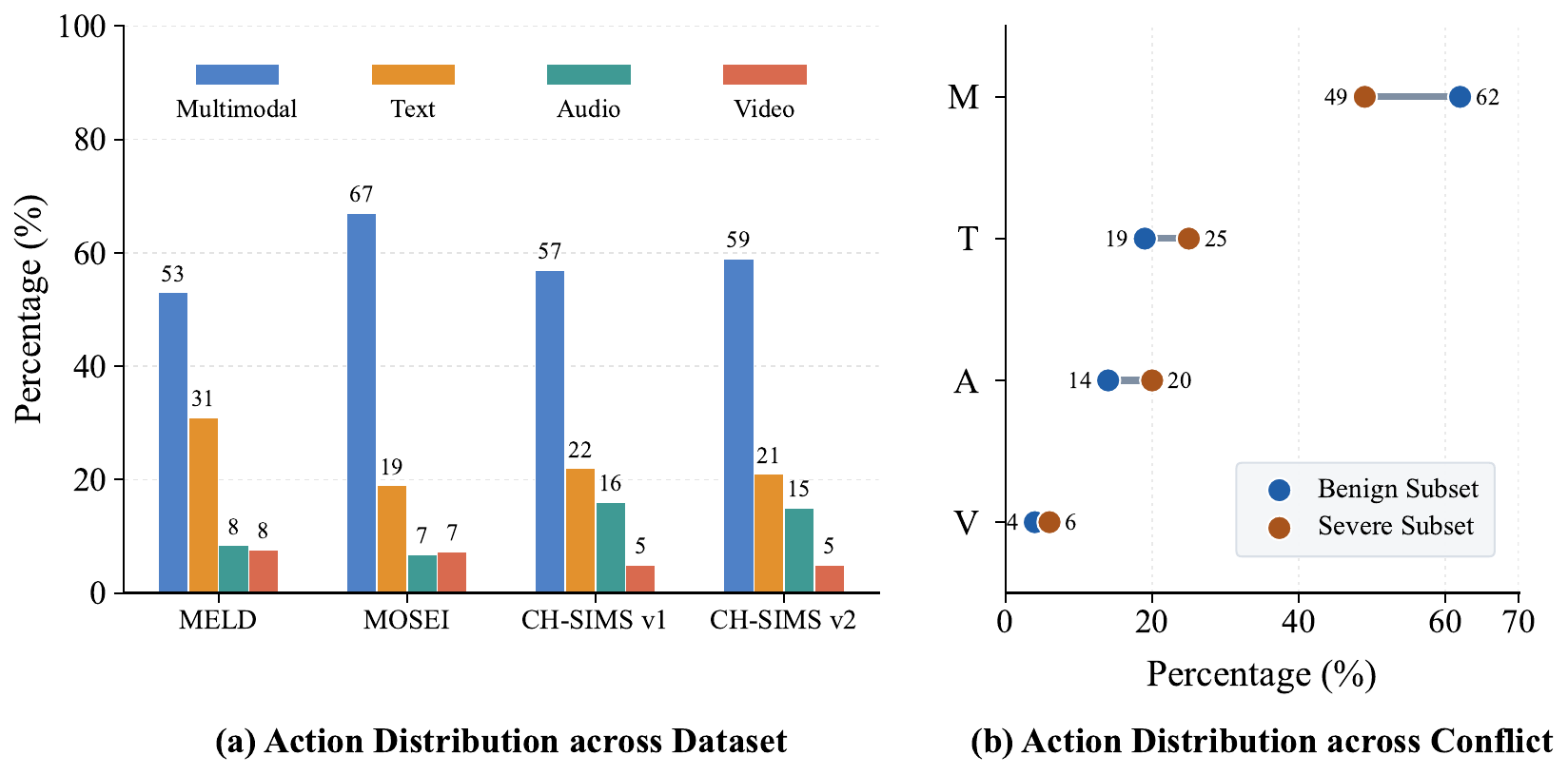} % 自动适配单栏宽度
  \vspace{-0.2cm}
  \caption{Modality selection distribution of ADA across datasets and conflict types. 
(a) The stacked bars show the selection frequency of each action path on MELD, CMU-MOSEI, CH-SIMS, and CH-SIMS v2, where different colors correspond to multimodal, textual, acoustic, and visual paths. 
(b) The horizontal comparison shows how ADA changes its selection preference between benign and severe conflict subsets. 
M, T, A, and V denote multimodal, textual, acoustic, and visual paths, respectively.}
  \vspace{-0.2cm}
  \label{fig:action_dis}
\end{figure}

\subsection{Distribution of Modality Selection in ADA}

To better understand the decision behavior of ADA, we analyze its modality selection distribution across different benchmarks and conflict scenarios. As shown in Figure~\ref{fig:action_dis}, the multimodal pathway is selected most frequently overall, followed by the textual modality, while the audio and visual pathways are chosen less often. This distribution provides insight into how the agent prioritizes different inference paths under modality inconsistency.

\vspace{0.5em}
\noindent \textbf{Cross-dataset Analysis.}
As shown in Figure~\ref{fig:action_dis}(a), the multimodal path remains the dominant choice across all four benchmarks, suggesting that ADA still prefers fusion when the modalities are sufficiently consistent. Meanwhile, the textual path is selected more frequently than the audio and visual ones, indicating that text often provides a more informative cue when cross-modal disagreement becomes harder to resolve. This trend is particularly evident on MELD, where the textual path is selected more often than on the other benchmarks, likely because richer dialogue context provides stronger semantic evidence for emotion inference.

\vspace{0.5em}
\noindent \textbf{Cross-conflict Type Analysis.}
We further examine ADA's behavior under different conflict types, as shown in Figure~\ref{fig:action_dis}(b). For benign conflicts, the agent relies primarily on the multimodal pathway (62\%), indicating that information integration is still beneficial when modality disagreement remains reconcilable. In contrast, under severe conflicts, the selection frequency of the multimodal path decreases to 49\%, while the reliance on unimodal paths increases, especially for the textual (25\%) and audio (20\%) branches. This shift indicates that ADA can adaptively reduce its dependence on fusion when modality contradictions become more pronounced, and instead route the prediction through more reliable unimodal cues.

\begin{table}[H]
    \centering
    \captionsetup{font=small}
    \small
    \setlength{\tabcolsep}{5pt}
    \caption{Accuracy comparison on benign and severe conflict subsets. \textbf{Bold} indicates the best result.}
    \label{tab:conflict_performance}
    \begin{tabular}{lcc}
        \specialrule{1.1pt}{0pt}{2pt}
        Methods & Benign Conflict & Severe Conflict \\
        \midrule
        MMML~\cite{wu-etal-2024-multimodal} 
        & 59.2\% & 40.2\% \\
        TelME~\cite{yun-etal-2024-telme} 
        & 61.8\% & 41.5\% \\
        FacialMMT~\cite{zheng-etal-2023-facial} 
        & 58.9\% & 44.1\% \\
        \midrule
        \rowcolor{lightblue}
        \textbf{DCR (ours)}
        & \textbf{72.4\%} & \textbf{50.3\%} \\
        \specialrule{1.1pt}{2pt}{3pt}
    \end{tabular}
\end{table}

\subsection{Performance on Conflict Subsets}
% We further evaluate the robustness of DCR against conventional fusion models on the benign and severe conflict subsets of CH-SIMS~\cite{yu-etal-2020-ch}, which are partitioned according to the heuristic conflict definition introduced in Section~\ref{sec:method}. As shown in Table~\ref{tab:conflict_performance}, DCR achieves clear accuracy improvements on both subsets. Specifically, it reaches 72.4\% accuracy on benign conflicts and 50.3\% on severe conflicts, substantially outperforming representative baselines such as TelME, which attains 61.8\% and 41.5\%, respectively.
We further evaluate DCR on conflict-specific subsets derived from CH-SIMS~\cite{yu-etal-2020-ch}. We use CH-SIMS here because its multimodal annotations allow us to construct conflict subsets following the heuristic definition in Section~\ref{sec:def_con}. This results in 308 benign-conflict samples and 149 severe-conflict samples for evaluation.
As shown in Table~\ref{tab:conflict_performance}, DCR achieves clear accuracy improvements on both subsets. Specifically, it reaches 72.4\% accuracy on benign conflicts and 50.3\% on severe conflicts, substantially outperforming representative baselines such as TelME, which attains 61.8\% and 41.5\%, respectively. Although the improvement is more pronounced on benign conflicts, DCR also maintains a consistent advantage on severe conflicts, where modality contradictions are stronger and conventional fusion is more prone to failure. These results indicate that DCR can effectively handle different degrees of modality conflict and remains robust under both reconcilable and irreconcilable disagreement scenarios.

\subsection{Exploration of Expanded Action Space}
\label{sec:exploration of expanded action space}
\begin{figure}[!t]
  \centering
  \includegraphics[width=\columnwidth]{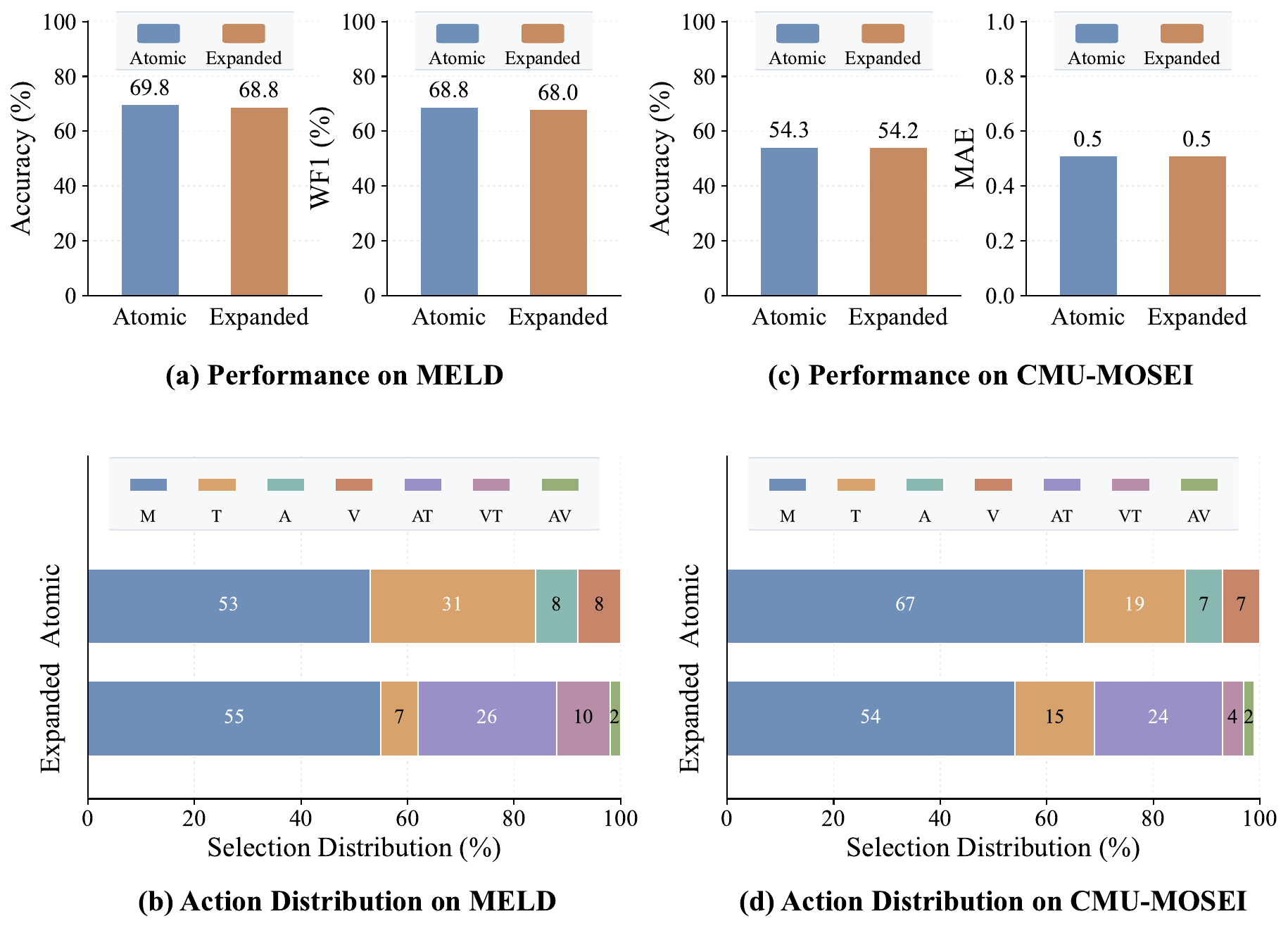} % 自动适配单栏宽度
  \vspace{-0.2cm}
  \caption{Comparison of atomic and expanded action spaces on MELD and CMU-MOSEI. (a) and (c) report the performance under the two action space designs, while (b) and (d) show the corresponding action selection distributions. Here, M, T, A, and V represent the original atomic actions, and AT, VT, and AV represent the additional pairwise actions in the expanded action space.}
  \vspace{-0.2cm}
  \label{fig:action_space}
\end{figure}

In ADA, we primarily adopt an atomic action space, where the agent selects either a single modality or the full multimodal representation. To examine whether a finer-grained routing strategy could further improve performance, we additionally explore an expanded action space that includes pairwise modality combinations, such as Text-Audio, Text-Video, and Audio-Video. The intuition is that these intermediate choices may help filter noise from specific modalities more flexibly.
However, as shown in Figure~\ref{fig:action_space}, introducing such pairwise actions does not improve performance on either MELD or CMU-MOSEI. Instead, we observe a slight decline in the overall results, while the agent still overwhelmingly prefers the original atomic actions even when additional pairwise options are available.

This result suggests that the original atomic action space already provides a sufficiently effective decision granularity for the current task. One possible reason is that expanding the action space increases the difficulty of policy learning, making optimization less stable under limited training data. In addition, the multimodal and unimodal experts in DCR already capture the major interaction patterns required for affective inference, so explicitly adding pairwise routes introduces limited new information while increasing structural redundancy. As a result, the expanded action space brings extra complexity without yielding consistent performance gains. Therefore, we retain the original four-way atomic action space as the default design in ADA.

\subsection{Additional Analysis on LLM-based Methods}
\label{sec:additional_analysis_iemocap_llm}

To provide a broader perspective on the generalization ability of DCR, 
we further compare DCR with representative LLM-based methods, including EmoCaps~\cite{li-etal-2022-emocaps}, BiosERC~\cite{Xue2024BiosERCIB}, InstructERC~\cite{lei2023instructerc}, and DialogueMMT~\cite{he-etal-2025-dialoguemmt}, on IEMOCAP and MELD. As reported in Table~\ref{tab:llm_based}, WF1 is used as the evaluation metric. Although DCR does not consistently outperform these LLM-based methods, it still achieves competitive results on both benchmarks. This performance gap is understandable, because dialogue-level emotion recognition relies heavily on contextual semantics, pragmatic cues, and implicit affective reasoning. LLM-based methods benefit from large-scale language pre-training and stronger textual understanding, which gives them an inherent advantage in modeling dialogue context and inferring emotions from complex linguistic expressions.

By contrast, DCR is not designed to compete with LLM-based methods through model scale or textual reasoning capacity. Instead, it focuses on a different and complementary problem: how to recognize and resolve modality conflicts when textual, acoustic, and visual cues provide inconsistent affective evidence. Therefore, the comparison does not weaken the motivation of conflict-aware modeling. Rather, it suggests that the proposed view of modality conflict remains meaningful, while stronger language backbones may further enhance the semantic understanding component of DCR. In this sense, DCR provides a lightweight and interpretable framework for conflict-aware multimodal emotion recognition, and its competitive performance indicates that explicit conflict resolution can bring useful gains even without relying on massive pretrained language models.

\begin{table}[!t]
    \centering
    \captionsetup{font=small}
    \small
    \setlength{\tabcolsep}{7pt}
    \caption{Comparison of WF1 scores with LLM-based methods on IEMOCAP and MELD. \textbf{Bold} indicates the best result.}
    \label{tab:llm_based}
    \begin{tabular}{lcc}
        \specialrule{1.1pt}{0pt}{2pt}
        Methods & IEMOCAP & MELD \\
        \midrule
        EmoCaps~\cite{li-etal-2022-emocaps} 
        & 71.11 & 64.00 \\
        BiosERC~\cite{Xue2024BiosERCIB} 
        & 69.02 & 68.72 \\
        InstructERC~\cite{lei2023instructerc} 
        & 71.39 & 69.15 \\
        \text{DialogueMMT}~\cite{he-etal-2025-dialoguemmt} 
        & \textbf{72.71} & \textbf{70.66} \\
        \midrule
        \rowcolor{lightblue}
        DCR (ours) 
        & 69.50 & 68.84 \\
        \specialrule{1.1pt}{2pt}{3pt}
    \end{tabular}
\end{table}

\begin{figure}[H]
  \centering
  \includegraphics[width=\columnwidth]{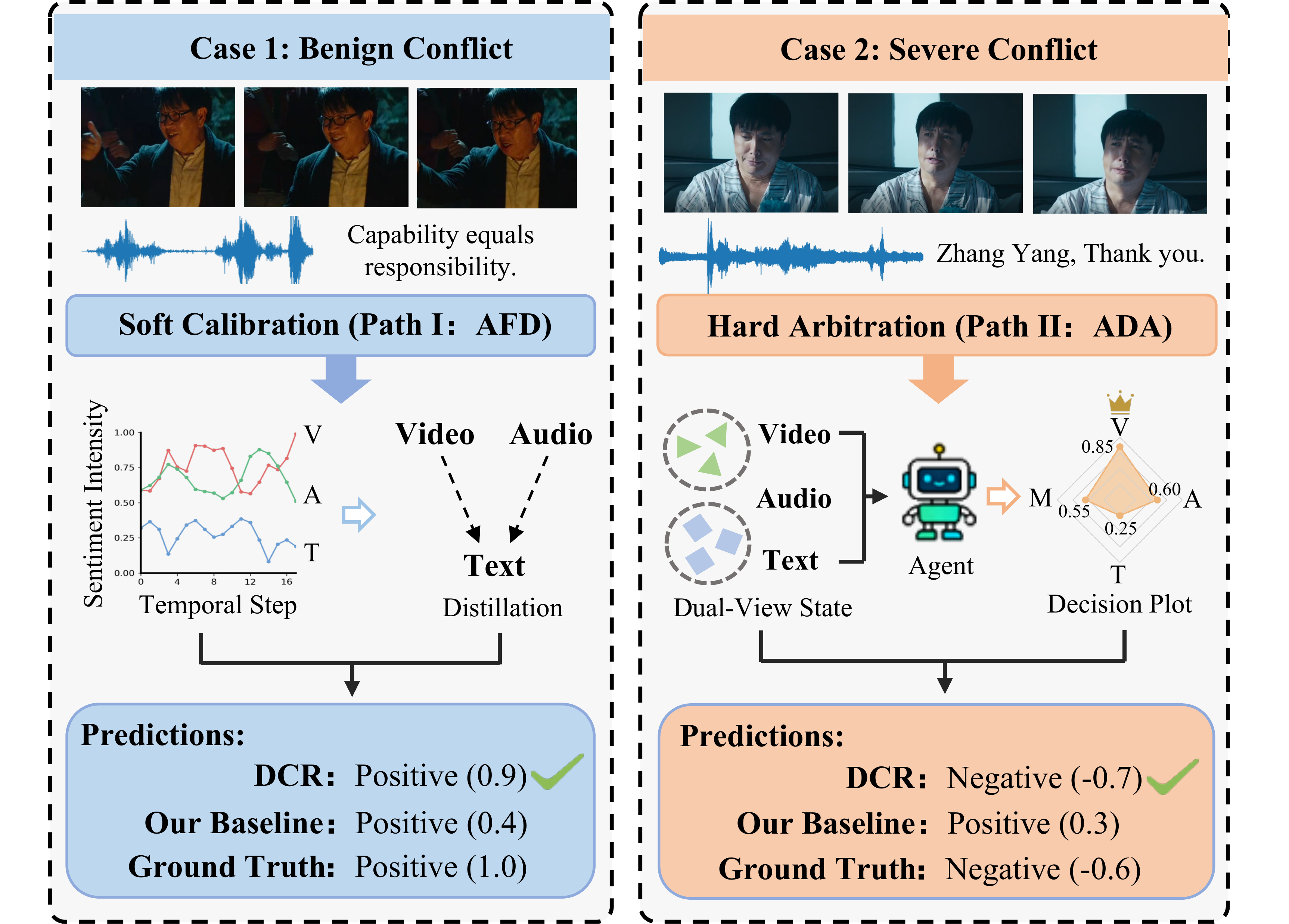}
  \vspace{-0.1cm}
  \caption{Illustration of DCR's conflict resolution logic. Case 1 demonstrates AFD-based soft calibration for reconcilable discrepancies, while Case 2 showcases ADA-based hard arbitration for severe, misleading conflicts.}
  \vspace{-0.2cm}
  \label{fig:case_study}
\end{figure}

\subsection{Case Study}
To qualitatively validate the efficacy of DCR and provide a granular view of its decision-making logic, we analyze representative cases of modality conflicts (see Figure~\ref{fig:case_study}). This section demonstrates how our framework adaptively navigates varying conflict intensities through active discernment, thereby enhancing the overall reliability of emotion recognition.

The efficacy of DCR stems from the hierarchical synergy between AFD and ADA, enabling the model to transition from soft calibration to hard arbitration based on signal resolvability. In Case 1 (\textit{“Capability equals responsibility.”}), the utterance presents a benign conflict where textual semantics alone lack sufficient resolution to capture the full sentiment intensity. AFD reconciles this discrepancy by distilling fine-grained temporal dynamics from non-verbal teachers into the textual student, yielding a high-confidence prediction (0.9) that aligns with the ground truth (1.0). Conversely, Case 2 (\textit{“Zhang Yang, Thank you.”}) exemplifies a sarcastic severe conflict where linguistic cues are intrinsically misleading. ADA addresses this by ``fact-checking'' subjective features against objective context, effectively suppressing the deceptive textual stream in favor of more credible visual and acoustic pathways. This strategic selection yields an accurate negative prediction (-0.7). Ultimately, these findings suggest that DCR's active discernment is indispensable for navigating the structural ambiguities and emotional complexities of human expression, promoting robust and credible inference across diverse scenarios.

\section{Conclusion}

% In this paper, we challenge the long-standing convention of passive multimodal aggregation by introducing the Dual-Path Conflict Resolution (DCR) framework. Through extensive evaluations across four diverse benchmarks, we demonstrate that recasting multimodal emotion recognition as an adaptive decision-making task significantly enhances resilience against cross-modal contradictions. Our findings reveal that robust affective analysis requires a sophisticated interplay between harmonizing complementary nuances and adjudicating polarized discrepancies. By shifting the focus from indiscriminate fusion to active discernment, this work establishes a new paradigm for building affective computing systems capable of navigating the inherent structural ambiguities and emotional complexities of human expression.

In this paper, we challenge the long-standing convention of passive multimodal aggregation by introducing the Dual-Path Conflict Resolution (DCR) framework. We further show that modality conflicts in MER differ in their resolvability, and should be treated accordingly. Through extensive evaluations across five diverse benchmarks, we demonstrate that recasting multimodal emotion recognition as an adaptive decision-making task significantly enhances resilience against cross-modal contradictions. Our findings reveal that robust affective analysis requires a sophisticated interplay between harmonizing complementary nuances and adjudicating polarized discrepancies, which are respectively handled by the two paths in DCR. By shifting the focus from indiscriminate fusion to active and conflict-aware discernment, this work establishes a new paradigm for building affective computing systems capable of navigating the inherent structural ambiguities and emotional complexities of human expression.

\bibliographystyle{IEEEtran}
\bibliography{reference}

\end{document}